\documentclass[a4paper,11pt]{article}
\pdfoutput=0 

\usepackage{jheppub} 
\usepackage{slashed}

\usepackage{amsmath,amssymb}
\usepackage{mathtools}
\usepackage{amsthm}
\usepackage{graphicx}

\newcommand{\bfm}[1]{\mbox{\boldmath$#1$}}

\newcommand{\gsim}{\;\rlap{\lower 3.5 pt \hbox{$\mathchar \sim$}} \raise 1pt
\hbox {$>$}\;}
\newcommand{\lsim}{\;\rlap{\lower 3.5 pt \hbox{$\mathchar \sim$}} \raise 1pt
\hbox {$<$}\;}

\newcommand{\Dplus}[1]{\left[\frac{\log^{#1}(1-z)}{1-z}\right]_+}
\newcommand{\DplusOne}[0]{\left[\frac{\log(1-z)}{1-z}\right]_+}
\newcommand{\DplusZero}[0]{\left[\frac{1}{1-z}\right]_+}

\title{\boldmath Light Quark Mediated  Higgs Boson Threshold Production
in the Next-to-Leading Logarithmic Approximation}

\preprint{ALBERTA-THY-02-20}

\author[a]{Charalampos Anastasiou,}
\author[b]{Alexander Penin}

\affiliation[b]{Institute for Theoretical Physics, ETH Z\"urich, 8093 Z\"urich, Switzerland}
\affiliation[a]{Department of Physics, University of Alberta, Edmonton AB T6G
2J1, Canada}

\emailAdd{babis@phys.ethz.ch}
\emailAdd{penin@ualberta.ca}

\abstract{We study the amplitude of the Higgs boson  production in gluon
fusion mediated by a light quark loop and evaluate the logarithmically
enhanced radiative corrections to the next-to-leading logarithmic
approximation which sums up  the terms of the form
$\alpha_s^n\ln^{2n-1}(m_H/m_q)$ to all orders in the strong coupling
constant. This result is used for the calculation of the process cross
section  near the production threshold and gives a quantitative estimate of
the three and four-loop   bottom quark contribution to the Higgs boson
production at the Large Hadron Collider.}

\begin{document}
\maketitle
\flushbottom

\section{Introduction}

Accurate theoretical predictions for the (inclusive and differential) Higgs
gluon fusion cross section are indispensable for the determination with high
precision of the Higgs boson couplings~\cite{Cepeda:2019klc}. A dominant
component of the gluon fusion process originates from Feynman diagrams with a
virtual top quark  inside the loop.  Due to the hierarchy of the top quark and
Higgs boson masses, this component can be accurately determined by expanding
around the heavy top quark
limit~\cite{Harlander:2009my,Pak:2009dg,Harlander:2009mq}. In this approach,
where top quark loops are reduced to effective point-like vertices, gluon fusion
cross sections are now known precisely at very high orders in perurbative
QCD~\cite{Anastasiou:2015ema,Anastasiou:2016cez,Mistlberger:2018etf,
Banfi:2015pju,Chen:2016zka,Caola:2015wna,Chen:2014gva,Dulat:2017prg,
Dulat:2018bfe,Cieri:2018oms}.

With the achieved precision of a few percent, contributions of lighter quarks of
a suppressed Higgs Yukawa coupling cannot be ignored.\footnote{For a recent
estimate of their effect to the inclusive Higgs cross-section see, for example,
Ref.~\cite{Anastasiou:2016cez}.} For light quarks, the top quark effective field
theory calculations are inapplicable. The relevant Higgs production probability
amplitudes need to be computed with their exact quark mass dependence or,
alternatively, by means of a systematic expansion  around the antithetic
asymptotic limit in which the quark mass is vanishing. The exact quark mass
dependence for the $gg \to H$ amplitude is only known through two
loops~\cite{Graudenz:1992pv,Djouadi:1991tka,Spira:1995rr,Harlander:2005rq,
Aglietti:2006tp,Bonciani:2007ex,Anastasiou:2006hc,Anastasiou:2009kn}. The
two-loop amplitudes for the top-bottom interference in the next-to-leading order
cross section \cite{Melnikov:2016qoc,Melnikov:2017pgf,Lindert:2017pky} for the
production of a Higgs boson in association with a jet have been computed by
means  of a small quark mass expansion.

In the small quark mass limit the radiative corrections are enhanced   by a
power of the logarithm $\ln(m_H/m_q)$ of the Higgs boson to a light quark
mass ratio.   For the physical values of the bottom and charm quark masses
the numerical value of the logarithm is quite large and it is  important to
control the size of the logarithmic corrections beyond two loops. For the $gg
\to H$ amplitude the leading (double) logarithmic corrections have been
evaluated to all orders in  strong coupling constant $\alpha_s$   in
Refs.~\cite{Liu:2017vkm,Liu:2018czl}. The abelian part of the
double-logarithmic corrections  for the  $gg \to Hg$ amplitude of Higgs plus
jet production has been obtained in   Ref.~\cite{Melnikov:2016emg}.  Though
the leading logarithmic approximation gives a qualitative estimate of the QCD
corrections beyond two loops,  subleading logarithmic corrections can be
numerically important. Their computation is necessary to quantifying the
theoretical uncertainty estimate. In particular, in the leading logarithmic
approximation the numerical predictions vary significantly with values of
light quark masses being taken in different renormalization schemes. The
logarithmic terms sensitive to ultraviolet renormalization, which cancel the
dependence of the amplitude on the renormalization scheme, are  formally
subleading. Extending the analysis beyond the leading logarithmic
approximation is therefore highly desired. In this paper, we  present the
analysis  of the light quark loop mediated  $gg \to H$ amplitude in  the
next-to-leading logarithmic  approximation which sums up the terms  of the
form  $\alpha_s^n\ln^{2n-1}(m_H/m_q)$ to all orders in perturbation theory
and use this result for the evaluation of the bottom quark effect on the
Higgs boson production cross section near the threshold.

The paper is organized as follows. In the next section we discuss the general
structure of the leading and next-to-leading logarithmic approximation, derive
the factorization and perform the  resummation of the next-to-leading
logarithmic corrections to a model mass-suppressed amplitude of  quark
scattering by a scalar gauge field operator.  In Sect.~\ref{sec::3} we apply the
method to the analysis of the $gg \to Hg$ amplitude mediated by a light quark
loop.  The result is used  in Sect.~\ref{sec::4} for the calculation of the
bottom quark contribution  to the  cross section of the Higgs boson production
in  gluon fusion in the threshold approximation. Sect.~\ref{sec::5} is our
conclusion.

\section{General structure of the leading and next-to-leading logarithms}
\label{sec::2}

The logarithmically enhanced contributions under
consideration appear in the high-energy or small-mass limit
of the on-shell gauge theory amplitudes. To the leading
order of the small-mass expansion these are  renowned
``Sudakov'' logarithms which have been extensively studied
since the pionering work \cite{Sudakov:1954sw}.  The
structure of the Sudakov logarithms in the theories with
massive fermions and gauge bosons is by now well understood
\cite{Frenkel:1976bj,Smilga:1979uj,Mueller:1979ih,Collins:1980ih,Sen:1981sd,
Sterman:1986aj,Korchemsky:1988hd,Korchemsky:1988pn,Kuhn:1999nn,Kuhn:2001hz,Feucht:2004rp,
Jantzen:2005az,Penin:2005eh,Penin:2005kf,Bonciani:2007eh,Bonciani:2008ep,
Kuhn:2007ca,Kuhn:2011mh,Penin:2011aa}. The characteristic
feature of the $gg\to H$ amplitude however  is that it
vanishes in the limit of the massless quark  $m_q \to 0$
{\it i.e.} is power-suppressed. The asymptotic behavior of
the power suppressed amplitudes  may be  significantly
different from the Sudakov case and  attract  a lot of
attention  in many various contexts (see {\it e.g.}
\cite{Liu:2017vkm,Liu:2018czl,Melnikov:2016emg,
Gorshkov:1966ht,Kotsky:1997rq,Akhoury:2001mz,Ferroglia:2009ep,
Laenen:2010uz,Banfi:2013eda,Becher:2013iya,deFlorian:2014vta,Anastasiou:2014lda,
Penin:2014msa,Almasy:2015dyv,Penin:2016wiw,Bonocore:2016awd,
Boughezal:2016zws,Moult:2017rpl,Liu:2017axv,Beneke:2017ztn,Boughezal:2018mvf,
Bruser:2018jnc,Moult:2018jjd,Ebert:2018lzn,Alte:2018nbn,Beneke:2018rbh,
Beneke:2018gvs,Engel:2018fsb,Ebert:2018gsn,Penin:2019xql,Beneke:2019mua,Liu:2019oav,Wang:2019mym}).
At the same time the logarithmically enhanced corrections
to the on-shell (or almost on-shell) amplitudes in the high
energy limit are universally  associated with the emission
of the virtual particles which are soft and/or collinear to
the large external  momenta. It is instructive  first to
review the origin and the structure of such corrections to
the leading-power amplitudes. In the next section we
discuss the asymptotic behavior of the  quark form factor
to the next-to-leading logarithmic approximation, which
will be used for  the analysis of the Higgs production in
Sect.~\ref{sec::3}.

\subsection{Sudakov   form factor}
\label{sec::2.1}
We consider the limit of the large Euclidean momentum transfer
$Q^2=-(p_2-p_1)^2$  and slightly off-shell external Euclidean quark momenta
$m_q^2\ll-p_i^2\ll Q^2$. In this limit the ``Sudakov'' radiative  corrections
enhanced by the logarithm of the small ratio $p_i^2/Q^2$ are known to
exponentiate
\cite{Sudakov:1954sw,Frenkel:1976bj,Smilga:1979uj,Mueller:1979ih,Collins:1980ih,
Sen:1981sd} and to the next-to-leading logarithmic approximation  the Dirac
form factor $F_1$ of the quark scattering in an external abelian vector field
reads
\begin{eqnarray}
F_1^{NLL}&=&\exp\left\{-{C_F\alpha_s \over 2\pi}I_{DL}(p_1^2,p_2^2,0)
\left[1-\beta_0{\alpha_s \over 8\pi} \left(\ln\left({-p_1^2\over \mu^2}\right)
+\ln\left({-p_2^2\over\mu^2}\right)\right)\right]\right.
\nonumber\\
&+&\gamma_q^{(1)}{\alpha_s\over 2\pi}I_{SL}(p_1^2,p_2^2,0)\bigg\}\,,
\label{eq::F1}
\end{eqnarray}
where $\beta_0={11\over 3}C_A-{4\over 3}T_Fn_l$ is the one-loop beta-function
for $n_l$ light flavors, $C_F=(N_c^2-1)/(2N_c)$, $C_A=N_c$, $T_F={1\over 2}$
for the $SU(N_c)$ color group,  $\alpha_s\equiv\alpha_s(\mu)$ is the strong
coupling constant at the renormalization scale $\mu$, $\gamma_q^{(1)}=3C_F/2$
is the one-loop quark collinear anomalous dimension, and $I_{DL}$ ($I_{SL}$)
is the double (single) logarithmic one-loop virtual momentum integral. Let us
consider the evaluation of the above integrals in more detail. The
double-logarithmic contribution is generated by the  scalar three-point
integral
\begin{equation}
I_{DL}(p_1^2,p_2^2,m_q^2)={iQ^2\over \pi^2 }\int{{d^4l}\over l^2
\left((p_1-l)^2-m_q^2\right)\left((p_2-l)^2-m_q^2\right)}\,,
\label{eq::dlint}
\end{equation}
where $l$ is the gluon momentum. For $m_q^2=0$ the above integral can be
computed by using  the expansion by regions method
\cite{Beneke:1997zp,Smirnov:1997gx,Smirnov:2002pj} with the result
\begin{eqnarray}
I_{DL}(p_1^2,p^2_2,0)&=&\left[{1\over\varepsilon^2}-{1\over\varepsilon}
\left(\ln Q^2-\ln(-p_1^2)-\ln(-p_2^2)\right)+{1\over 2}\ln^2Q^2 \right.
\nonumber\\
&-&
\ln Q^2\left(\ln(-p_1^2)+\ln(-p_2^2)\right)
+\ln^2(-p_1^2)+\ln^2(-p_2^2)\bigg]_{us}
\nonumber\\
&+&\left[-{2\over\varepsilon^2}+{1\over\varepsilon}
\left(\ln(-p_1^2)+\ln(-p_2^2)\right)
+{1\over 2}\ln^2(-p_1^2)+{1\over 2}\ln^2(-p_2^2)\right]_c
\nonumber\\
&+&\left[{1\over\varepsilon^2}-{1\over\varepsilon}\ln Q^2+{1\over
2}\ln^2Q^2\right]_h+\ldots
\nonumber\\
&=&\ln
\left({Q^2\over -p_1^2}\right)\ln\left({Q^2\over -p_2^2}\right)+\ldots\,,
\label{eq::dlintreg}
\end{eqnarray}
where the ellipsis stand for the nonlogarithmic terms, $\varepsilon=(d-4)/2$ is
the parameter of dimensional regularization and the contributions of the
ultrasoft, collinear and hard regions are given separately (for application of
the expansion by regions to the form factor analysis see \cite{Kuhn:1999nn}).
Note that the logarithmic contribution can be read off the singularity
structure of the   ultrasoft and  collinear regions  where the quark propagators
may be  taken in the eikonal  approximation. As we see the integral $I_{DL}$
generates pure  double-logarithmic contribution associated with the overlapping
soft and collinear divergences and  does not generate any single logarithmic
term. On the other hand the double-logarithmic term can be obtained directly by
means of the original Sudakov method \cite{Sudakov:1954sw}. In this case the
propagators in Eq.~(\ref{eq::dlint})  are approximated  as follows
\begin{equation}
{1\over l^2}  \approx - i \pi \delta(Q^2uv - {l}_\perp^2)\,,\qquad
{1\over (p_1-l)^2}\approx -\frac{1}{Q^2v}\,,\qquad
{1\over (p_2 -l)^2} \approx -\frac{1}{Q^2u}\,,
\label{eq::prop}
\end{equation}
where we introduce  the standard Sudakov parametrization of
the  soft gluon momentum $l=up_1+vp_2+l_\perp$ with
Euclidean $l_\perp^2\ge 0$. The logarithmic scaling of the
integrand requires $-p_1^2/Q^2<|v|<1$,  $-p_2^2/Q^2<|u|<1$
and the additional kinematical constraints $uv>0$  has to
be imposed to ensure that the soft gluon propagator can go
on the mass shell. After integrating Eq.~(\ref{eq::dlint})
over ${l}_\perp$  we get
\begin{equation}
\int_{-p_1^2/Q^2}^{1}{{\rm d}v\over v}
\int_{-p_2^2/Q^2}^{1}{{\rm d}u\over u}=\ln
\left({Q^2\over -p_1^2}\right)\ln\left({Q^2\over -p_2^2}\right)\,.
\label{eq::dlintuv}
\end{equation}
Note that the  lower integration limits in  Eq.~(\ref{eq::dlintuv}) are
determined in the leading logarithmic approximation only up to a  constant
factor which we choose in such a way that  Eq.~(\ref{eq::dlintuv}) reproduces
the  result of the explicit evaluation Eq.~(\ref{eq::dlint}) to the
next-to-leading  logarithmic accuracy. As it follows from  Eq.~(\ref{eq::prop})
the logarithmic contribution to $I_{DL}$ is saturated by the ``soft'' virtual
momentum\footnote{This should not be confused with the  soft momentum region of
the expansion by regions approach.} with $l^2\approx 0$ corresponding to the
gluon propagator pole position  with the quark propagators carrying the large
external momenta being eikonal.

The single-logarithmic one-loop term gets contribution from the part of the
vertex diagram linear and quadratic in the virtual gluon momentum  as well as
from the quark self-energy diagram and can be reduced to the sum of two scalar
integrals
\begin{equation}
I_{SL}(p^2_1,p^2_2,m_q^2)={i\over 2\pi^2 }\int{{d^4l}\over (p_2-l)^2-m_q^2}
\left( {1\over l^2}-{1\over(p_1-l)^2-m_q^2}\right)+(p_1\leftrightarrow p_2)\,.
\label{eq::collint1}
\end{equation}
The first integral develops the  logarithmic contribution when the virtual
momentum becomes collinear to $p_1$. In the light-cone coordinates where
$p_1\approx p_1^-$ and $p_{2}\approx p_2^+$ it takes the following form
\begin{equation}
{i\over 2\pi }\int{{dl^+}{dl^-}{d{l}^2_\perp}\over (2p_2^+l^--l^2)}
\left( {1\over l^2}+{1\over (2p_1^-l^+-l^2)}\right)\,,
\label{eq::collint2}
\end{equation}
where we neglected $m_q^2$ and $p_i^2$. The integral over $l^-$ can then be
performed by taking the residue of the  quark propagator with the external
momentum  $p_2$ which gives
\begin{equation}
l^-={-{l}_\perp^2\over 2(p_2^+-l^+)}
\label{eq::residue}
\end{equation}
with a condition $0\le l^+\le p_2^+$   on the second
light-cone component of the virtual momentum. Then
Eq.~(\ref{eq::collint2}) takes the following form
\begin{equation}
\int_0^{p_2^+}{dl^+\over 2p_2^+}
\int{dl^2_\perp}\left( {1\over l_\perp^2}
-{1\over l_\perp^2+Q^2l^+(p_2^+-l^+)/(p_2^+)^2}\right)\,.
\label{eq::collint3}
\end{equation}
Since  $l^+\sim p^+$ the integral over the transversal
component of the virtual  momentum is logarithmic  for
${l}_\perp \ll Q$ and is regulated by $m_q^2$ or  $-p_2^2$
at small ${l}_\perp$. Thus for  $m_q^2\ll -p_1^2$ with the
logarithmic accuracy we get
\begin{equation}
\int_0^{p_2^+}{dl^+\over 2 p_2^+}\int^{Q^2}_{-p_2^2}
{dl^2_\perp \over l_\perp^2}={1\over 2}\ln\left({Q^2\over -p_1^2}\right)\,
\label{eq::collint4}
\end{equation}
and
\begin{equation}
I_{SL}(p^2_1,p^2_2,0)={1\over 2}\left[\ln\left({Q^2\over -p_1^2}\right)
+\ln\left({Q^2\over -p_2^2}\right)\right]+\ldots\,.
\label{eq::collintresult}
\end{equation}
The $\beta_0$ term in the exponent Eq.~(\ref{eq::F1}) sets the scale of the
strong coupling constant in the double-logarithmic contribution and is a
geometric average of the hard scale $Q^2$ and the ultrasoft scale $p_i^4/Q^2$,
as follows  from the  evolution equations  analysis
\cite{Korchemsky:1988hd,Kuhn:1999nn}.

For the analysis of the Higgs boson production we need a generalization of the
above result to the quark scalar form factor $F_S$. The structure of the Sudakov
logarithms  does not depend on the Lorenz structure of the amplitude. However,
in contrast to the  vector case the scalar form factor has a nonvanishing
anomalous dimension. The physical scale for the Yukawa coupling to an external
scalar field  is given by the momentum transfer which results in  an additional
dependence of the form factor on $Q$. Thus to the next-to-leading  logarithmic
accuracy we have
\begin{eqnarray}
F_S^{NLL}&=&\left({\alpha_s(Q)\over \alpha_s(\nu)}\right)^{\gamma_m^{(1)}/\beta_0}
\exp\left\{-{C_F\alpha_s \over 2\pi}\ln\left({Q^2\over -p_1^2}\right)
\ln\left({Q^2\over -p_2^2}\right)\left[1-\beta_0{\alpha_s \over 8\pi}
\left(\ln\left({-p_1^2\over \mu^2}\right)\right.\right.\right.
\nonumber\\
&+&\left.\left.
\ln\left({-p_2^2\over\mu^2}\right)\right)\right]
+\gamma_q^{(1)}{\alpha_s\over 4\pi}\left[\ln\left({Q^2\over -p_1^2}\right)
+\ln\left({Q^2\over -p_2^2}\right)\right]
\bigg\}\,,
\label{eq::FS}
\end{eqnarray}
where  the exponential factor is identical to Eq.~(\ref{eq::F1}), in the
renormalization  group factor  $\gamma^{(1)}_m={3}C_F$ is the quark mass
anomalous dimension, and $\nu$ is the  renormalization scale of the Yukawa
coupling.

The light quark mediated $gg\to H$ amplitude is suppressed by the quark mass and
its high-energy asymptotic behavior is significantly different from the
leading-power Sudakov form factor considered above. To  determine  the structure
of the logarithmically enhanced corrections in this case in the next section we
consider   an auxiliary amplitude  of a massive quark scattering by an abelian
gauge field operator. This rather artificial amplitude  is a perfect toy model
which  reveals the main features of the problem in the most illustrative way
with minimal technical complications and the corresponding analysis can be
easily generalized to the other mass-suppressed amplitudes and nonabelian gauge
groups.

\subsection{Massive quark scattering by a gauge field operator}
\label{sec::2.2}
\begin{figure}
\begin{center}
\begin{tabular}{ccc}
\includegraphics[width=1.5cm]{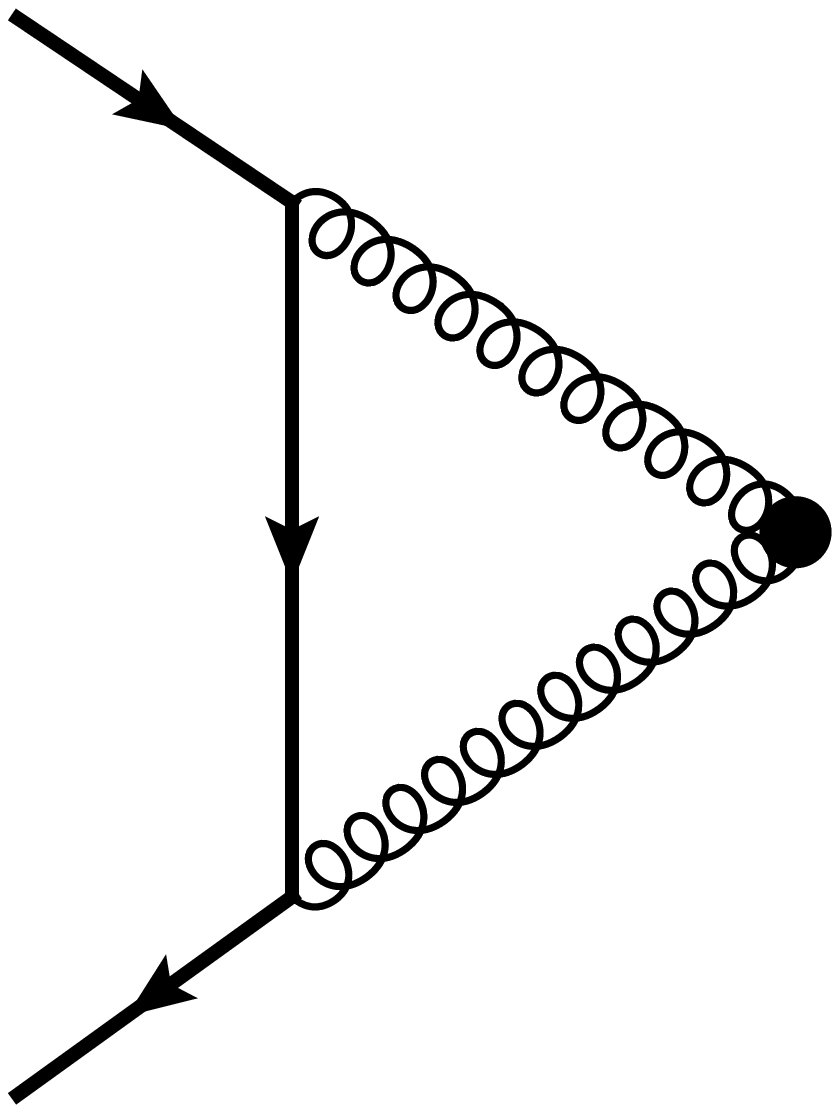}&
\hspace*{03mm}\includegraphics[width=1.5cm]{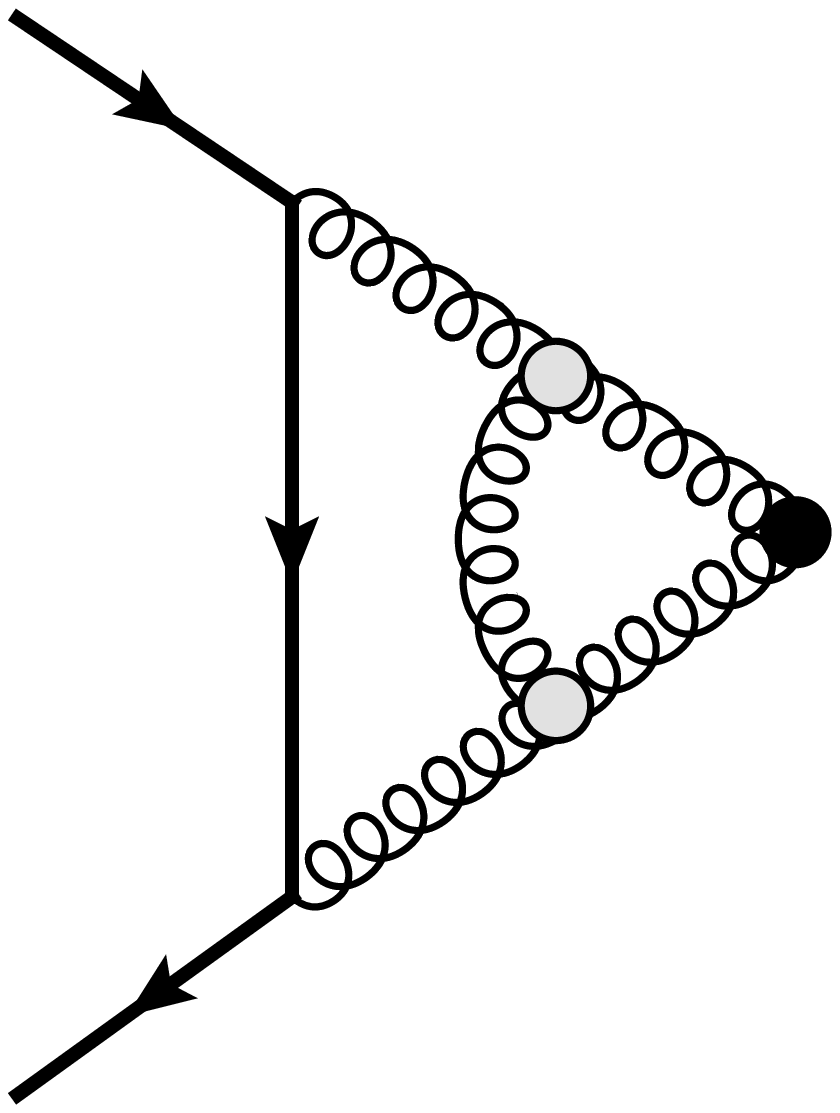}&
\hspace*{03mm}\includegraphics[width=1.5cm]{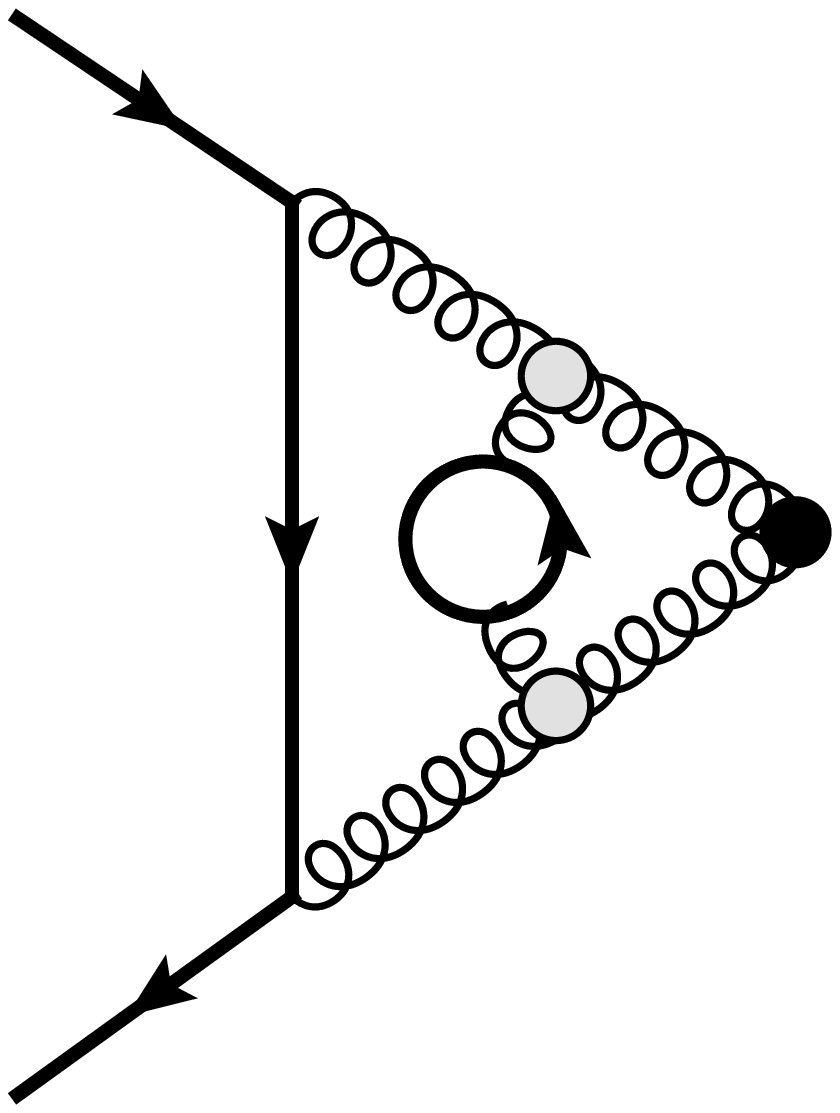}\\
(a)&\hspace*{03mm}(b)&\hspace*{03mm}(c)\\
\end{tabular}
\end{center}
\caption{\label{fig::1}  The Feynman diagrams for (a) the leading order one-loop
quark scattering by the $(G_{\mu\nu})^2$ vertex (black circle), (b) the soft
gauge boson exchange with the effective vertices (gray circles) defined in the
text which represents the non-Sudakov double-logarithmic corrections, (c) the
renormalization group running of the effective coupling constant in (b).}
\end{figure}

We consider  quark scattering by a local operator $(G_{\mu\nu})^2$ of an abelian
gauge field strength tensor for the  on-shell initial and final quark momentum
$p_1^2=p_2^2=m_q^2\ll Q^2$.   A detailed discussion of this process in the
leading logarithmic approximation can be found in
Refs.~\cite{Liu:2017vkm,Liu:2018czl}. Below we extend the analysis to the
next-to-leading logarithmic approximation. The leading order scattering is given
by the one-loop diagram in Fig.~\ref{fig::1}(a).  Due to helicity  conservation
the corresponding amplitude is suppressed    at high energy by the quark mass
and  with the logarithmic accuracy reads
\begin{equation}
{\cal G}^0= {\alpha_q\over \pi} \left(I'_{DL}(m_q^2,m_q^2,m_q^2)+
3I'_{SL}(m_q^2,m_q^2,m_q^2)- 3I_{UV}\right)m_q \,\bar{q}q\,,
\label{eq::G0}
\end{equation}
where, $\alpha_q=e_q^2/(4\pi)$,  $e_q$ is the quark abelian charge and the
scalar double-logarithmic integral over the virtual quark momentum reads
\begin{equation}
I'_{DL}(p_1^2,p_2^2,m_q^2)={iQ^2\over \pi^2 }\int{{d^4l}\over (l^2-m_q^2)
(p_1+l)^2(p_2 +l)^2}\,.
\label{eq::dlint'}
\end{equation}
The integral can be evaluated through the expansion by regions with the result
\begin{eqnarray}
I'_{DL}(m_q^2,m_q^2,m_q^2)&=&\left[-{2\over\varepsilon^2}+{1\over\varepsilon}
\ln(Q^2)-\ln m_q^2\ln^2Q^2+{1\over 2}\ln^2m_q^2\right]_c
\nonumber\\
&+&\left[{1\over\varepsilon^2}-{1\over\varepsilon}\ln Q^2+{1\over2}
\ln^2Q^2\right]_h+\ldots
\nonumber\\
&=&{L^2\over 2}+\ldots\,,
\label{eq::dlintregp}
\end{eqnarray}
where $L=\ln(Q^2/m_q^2)$ and the ellipsis stand for the nonlogarithmic terms.
The single-logarithmic collinear contribution is given by the integral
\begin{equation}
I'_{SL}(p^2_1,p^2_2,m_q^2)={i\over 2\pi^2 }\int{{d^4l}\over (p_2-l)^2}
\left( {1\over l^2-m_q^2}-{1\over(p_1-l)^2}\right)+(p_1\leftrightarrow p_2)
=L+\ldots\,,
\label{eq::collint'}
\end{equation}
and the ultraviolet  divergent contribution reads
\begin{equation}
I_{UV}={i\over \pi^2 }\int{{d^{4-2\varepsilon}l}\over (p_2-l)^2(p_1-l)^2}
=-{1\over \varepsilon}+\ln\left(Q^2\over \nu^2\right)+\ldots\,,
\label{eq::uvint}
\end{equation}
where $\nu$ is the corresponding ultraviolet renormalization scale. For
$\nu=m_q$ the linear logarithmic terms in Eq.~(\ref{eq::G0}) cancel and the
renormalized amplitude in the next-to-leading logarithmic approximation takes
a simple form
\begin{equation}
{\cal G}^0= {\alpha_qL^2\over 2\pi}m_q \,\bar{q}q\,.
\label{eq::G0ren}
\end{equation}
The integral $I'_{DL}$ in Eq.~(\ref{eq::dlint'}) can be evaluated by using
the Sudakov parametrization in the same way as it was done in the previous
section for $I_{DL}$. After integrating over $l_\perp$ we get
\cite{Liu:2018czl}
\begin{equation}
\int_{m_q^2/Q^2}^{1}{{\rm d}v\over v}\int_{m_q^2/vQ^2}^{1}{{\rm d}u\over u}
=L^2\int_0^1 {\rm d}\xi \int_{0}^{1-\xi}{\rm d}\eta
={L^2\over 2}\,,
\label{eq::dlintluvp}
\end{equation}
where the normalized logarithmic variables read $\eta=-\ln v/L $, $\xi=-\ln
u/L$.  As in Eq.~(\ref{eq::dlintuv}) we choose the integration limits in such a
way that  Eq.~(\ref{eq::dlintluvp}) reproduces the result of explicit evaluation
Eq.~(\ref{eq::dlintregp}) to the next-to-leading logarithmic accuracy. A
characteristic  feature of the mass-suppressed amplitude is that in contrast to
the Sudakov form factor  the double-logarithmic contribution is generated by a
soft quark exchange between the eikonal gauge boson lines. The additional power
of $m_q$ originates from the numerator of the virtual quark propagator which
effectively becomes scalar and therefore is sufficiently singular at small
momentum to  provide the double-logarithmic scaling of the one-loop  amplitude.

\begin{figure}
\begin{center}
\begin{tabular}{ccccc}
\includegraphics[width=1.5cm]{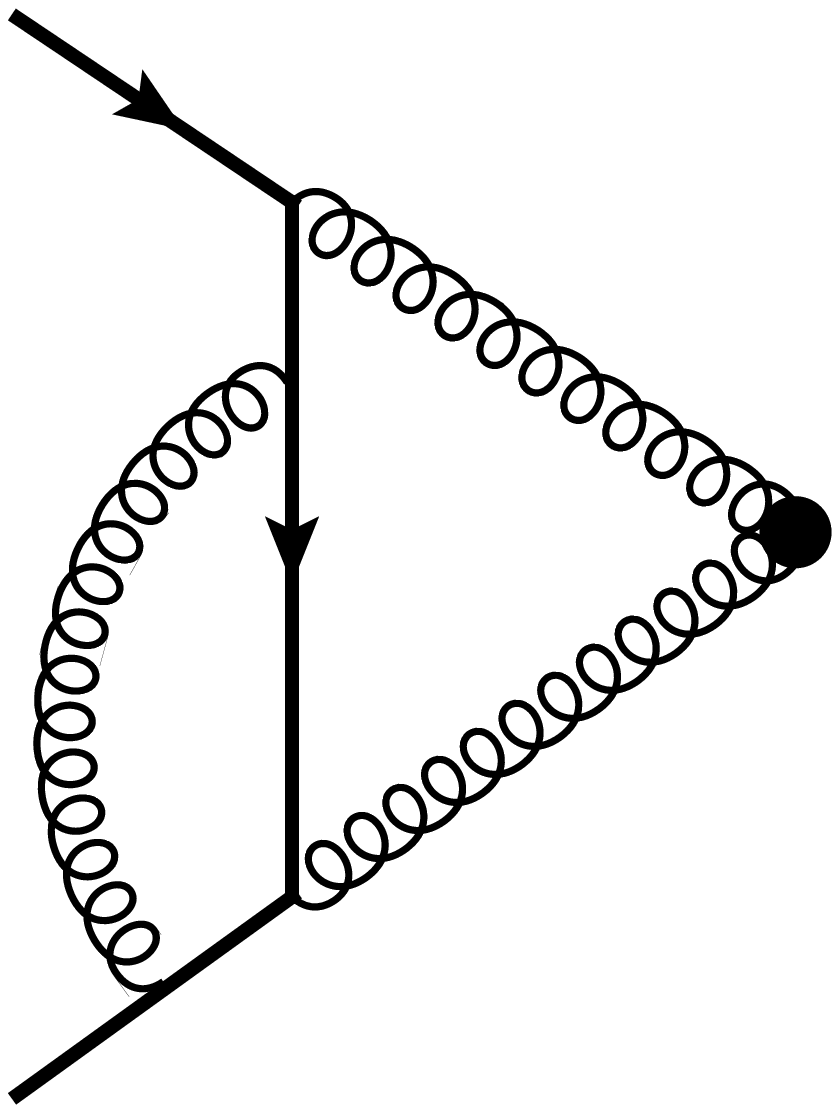}&
\hspace*{03mm}\includegraphics[width=1.5cm]{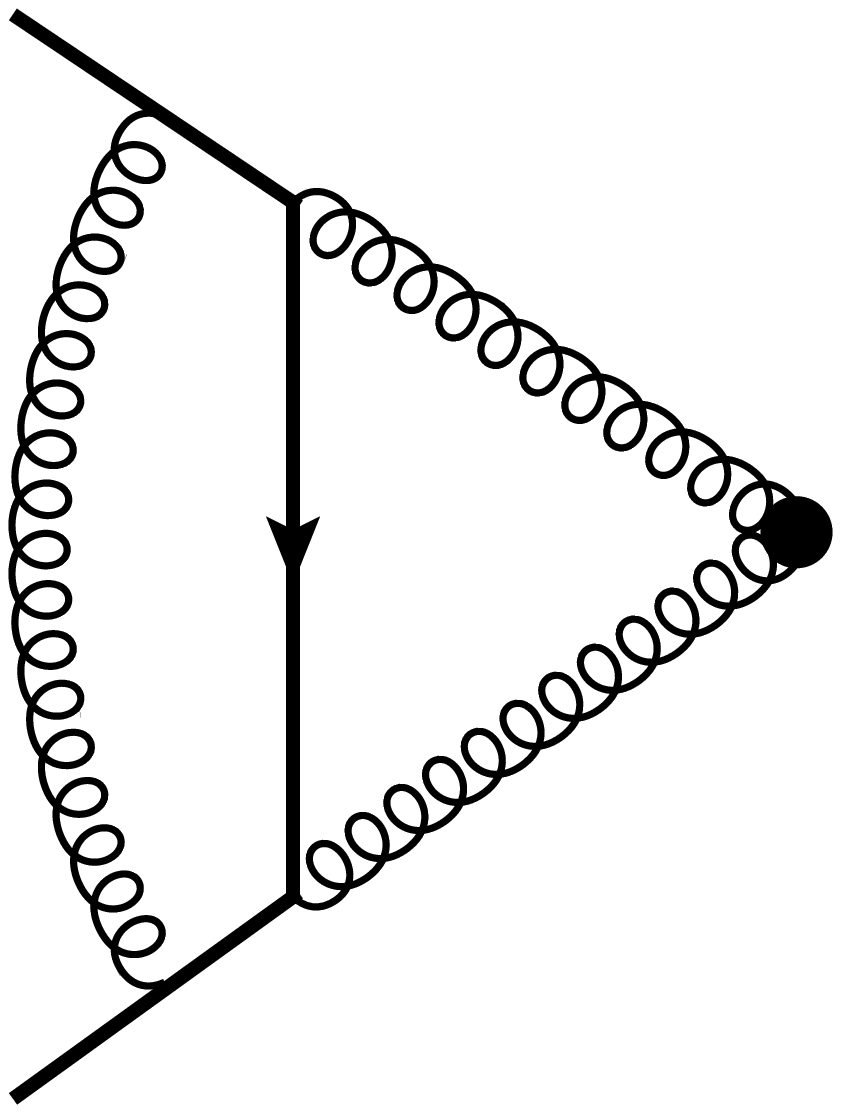}&
\hspace*{03mm}\includegraphics[width=1.5cm]{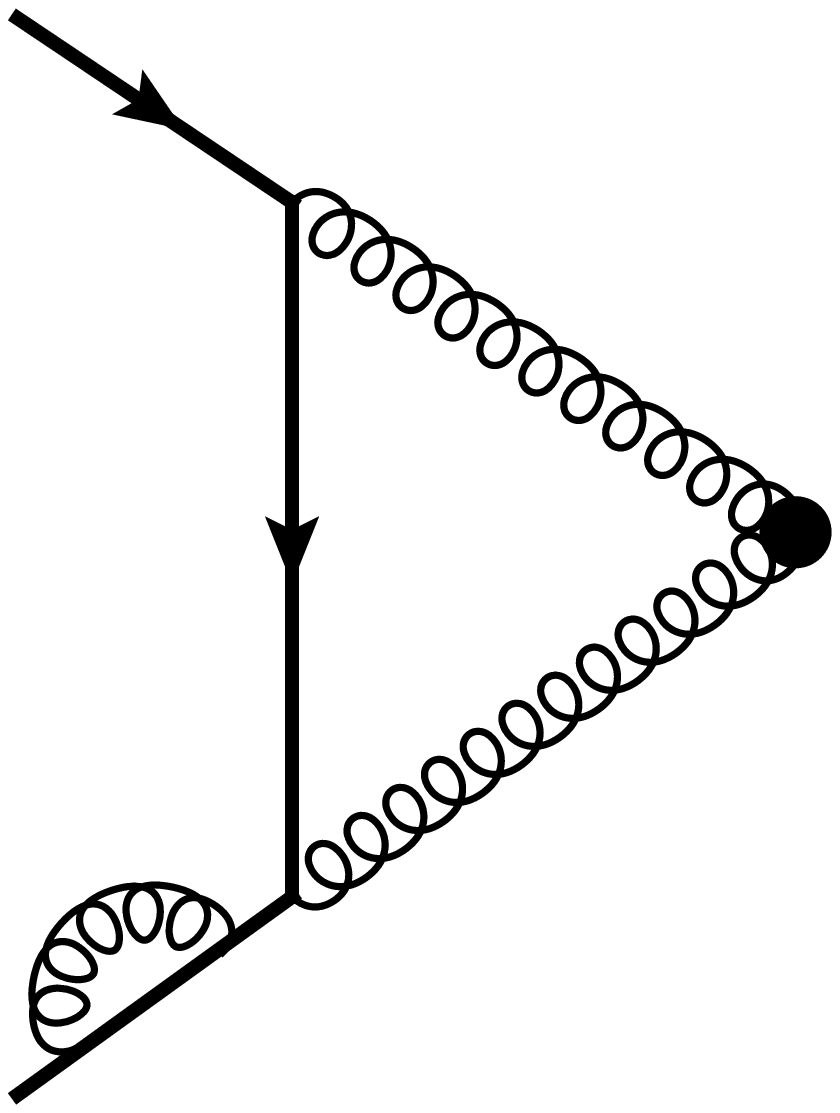}&
\hspace*{03mm}\includegraphics[width=1.5cm]{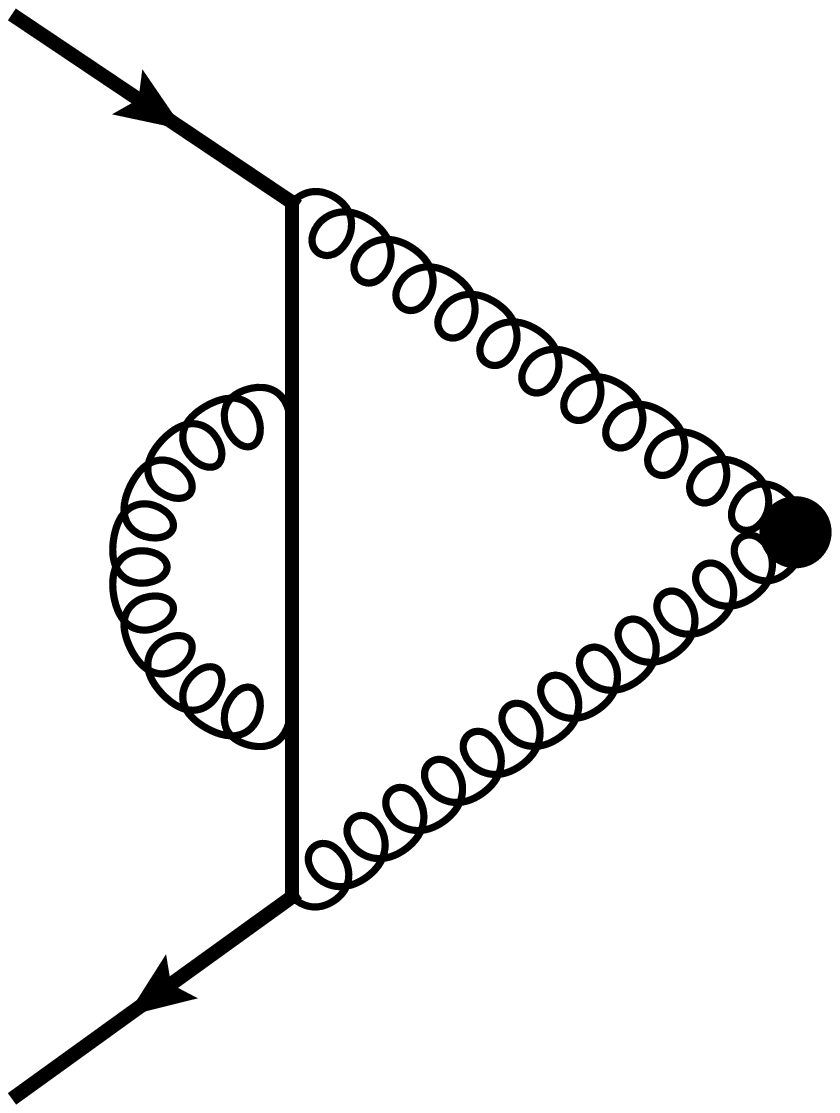}&
\hspace*{03mm}\includegraphics[width=1.5cm]{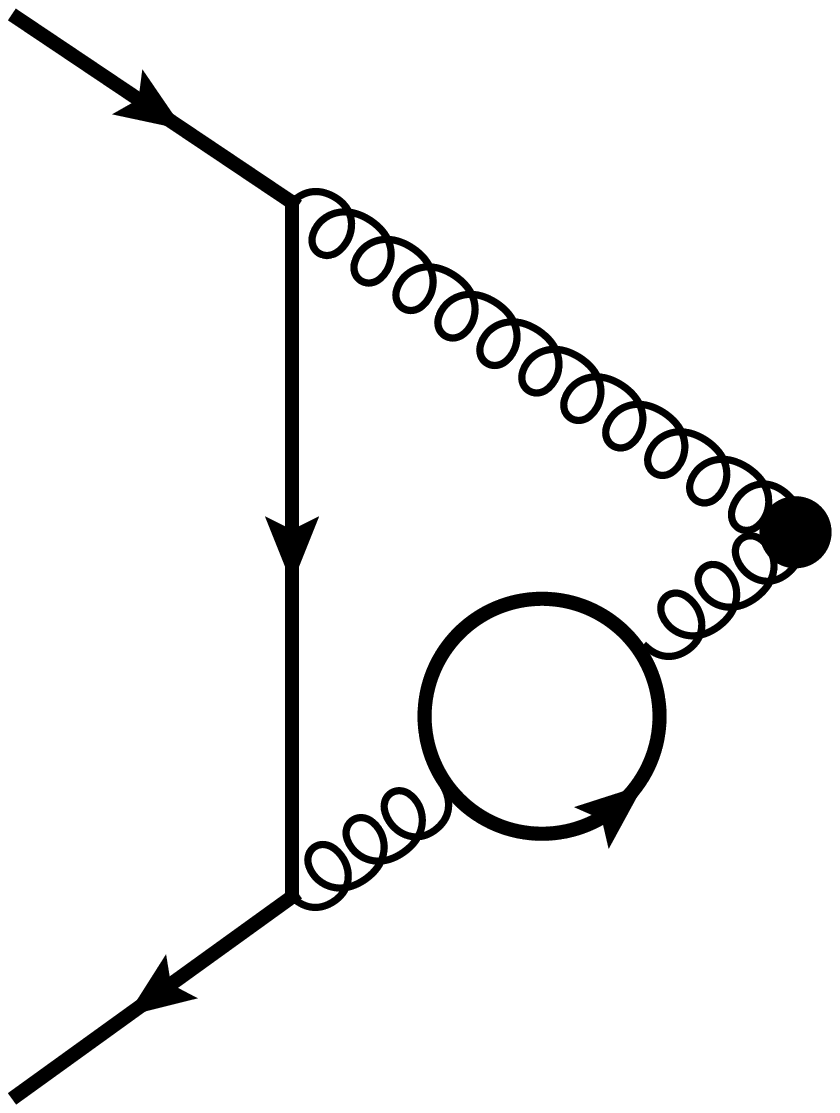}\\
(a)&\hspace*{03mm}(b)&\hspace*{03mm}(c)& \hspace*{03mm} (d)
& \hspace*{03mm} (e)\\
\end{tabular}
\end{center}
\caption{\label{fig::2} The two-loop  Feynman diagrams for the quark scattering
by the $(G_{\mu\nu})^2$ vertex (black circle) which contribute in the
next-to-leading logarithmic approximation. Symmetric diagrams are not shown.
}
\end{figure}

\begin{figure}
\begin{center}
\begin{tabular}{cccc}
\hspace{5mm}\raisebox{18.0mm}{\tiny $p_1$}\hspace{-4mm}
\raisebox{.5mm}{\tiny $p_2$}\hspace{-4mm}
\raisebox{7.0mm}{\tiny $l_g$}\hspace{-1mm}
\raisebox{0mm}{\includegraphics[width=1.7cm]{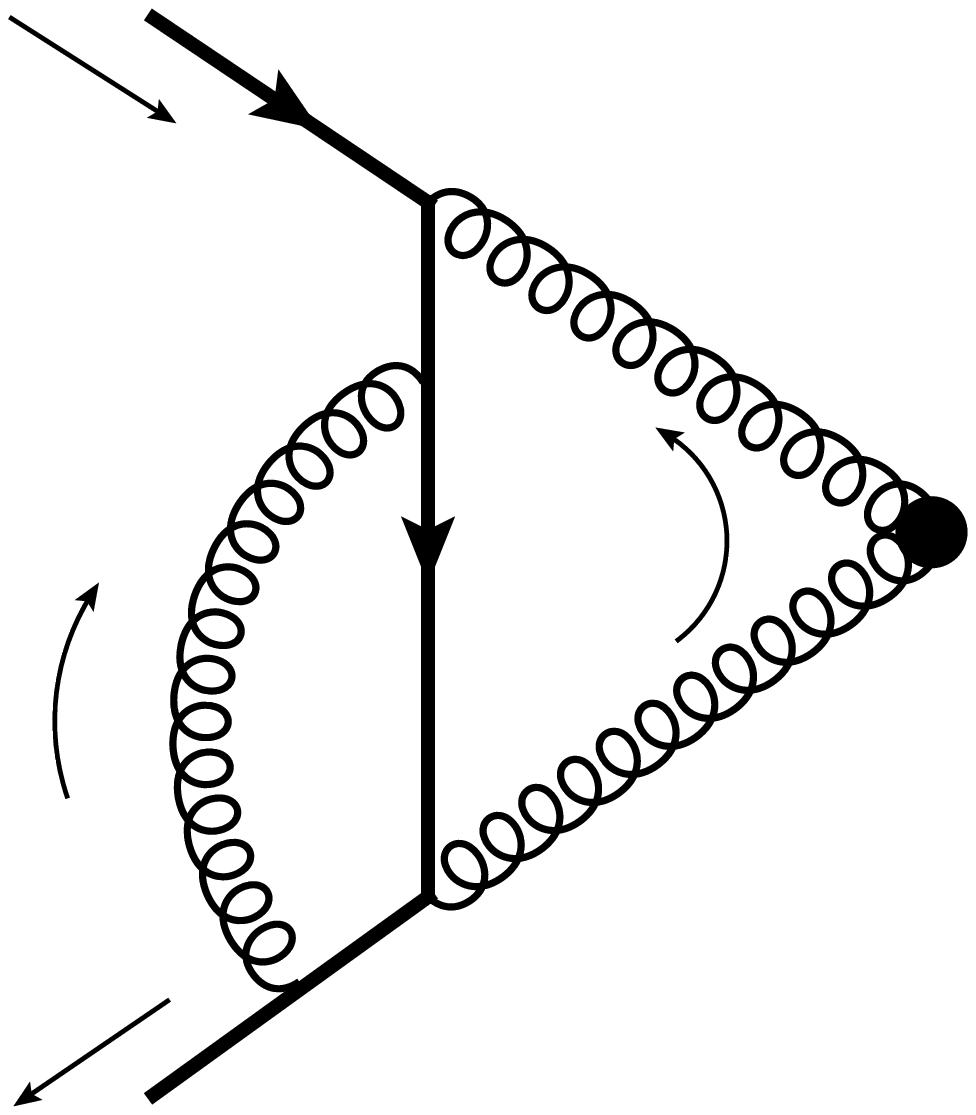}}
\hspace{-7.5mm}\raisebox{9.5mm}{\tiny $l$}
\hspace{6.0mm}~\raisebox{9.5mm}{$\bfm{\to}$}&
\hspace*{00mm}\includegraphics[width=1.5cm]{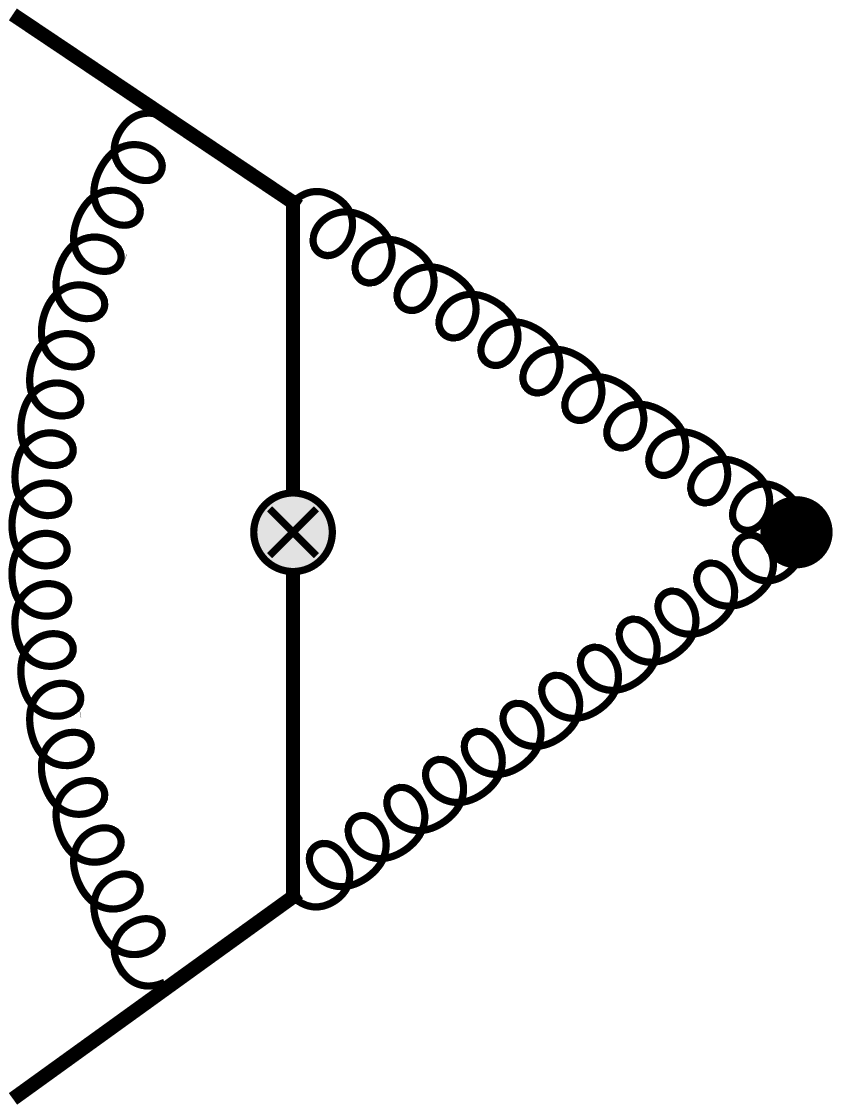}
~\raisebox{9.5mm}{$\bfm{\to}$}&
\hspace*{00mm}\includegraphics[width=1.5cm]{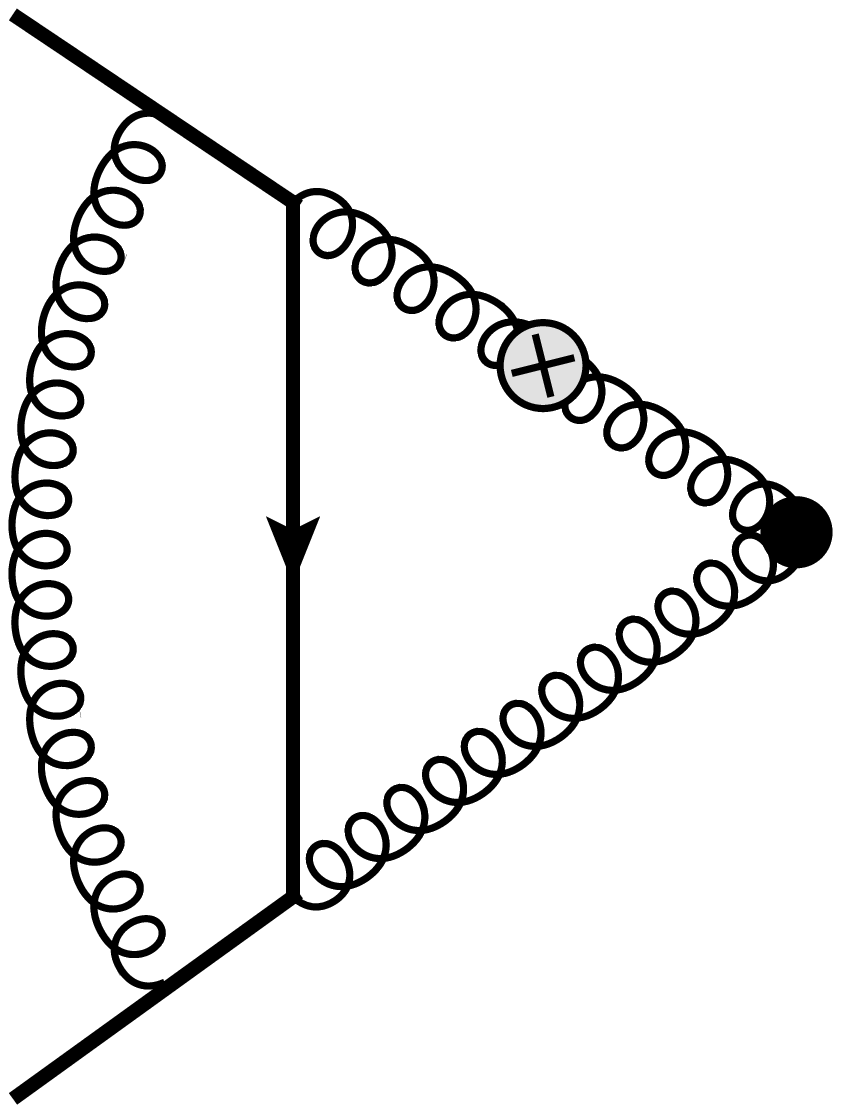}
~\raisebox{9.5mm}{$\bfm{\to}$}&
\hspace*{00mm}\includegraphics[width=1.5cm]{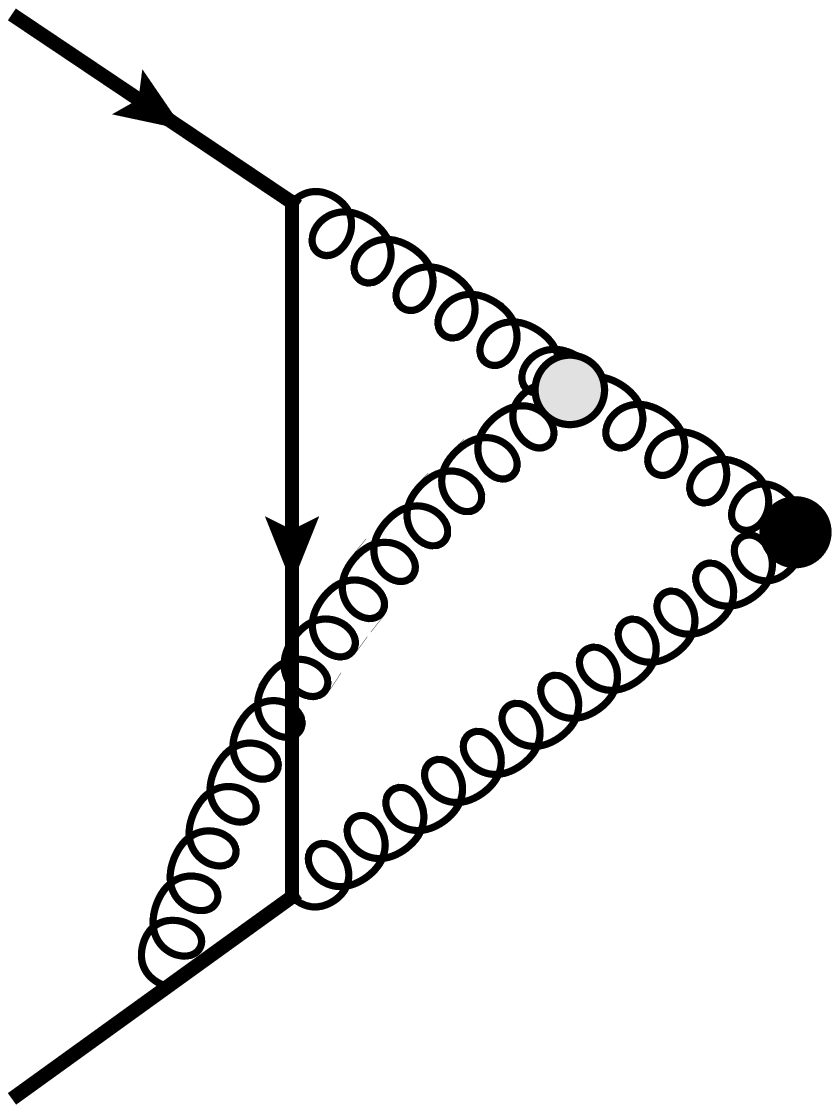}\\
(a) & \hspace*{-7mm}(b)
&\hspace*{-7mm}(c)& \hspace*{0.2mm} (d)
\\
\end{tabular}
\end{center}
\caption{\label{fig::3}  The diagram transformation which moves the
gauge boson vertex from the soft quark to the eikonal gauge boson line,
as explained in the text.}
\end{figure}

To determine the factorization structure of the next-to-leading logarithms we
consider  the two-loop radiative corrections and start with  the diagrams
Fig.~\ref{fig::2}(a,b), which include all the double-logarithmic
contributions. Following  Refs.~\cite{Liu:2017vkm,Liu:2018czl} we use  a
sequence of identities graphically represented in Fig.~\ref{fig::3}  to move
the gauge boson  vertex in the diagram Fig.~\ref{fig::2}(a) from the virtual
quark line  to an eikonal photon line. Let us describe this procedure in more
detail. We choose the virtual momentum routing as shown in
Fig.~\ref{fig::3}(a).  To the next-to-leading logarithmic approximation the
integration over at least one of the virtual momenta has to be
double-logarithmic. We start with the case when this is  the virtual  quark
momentum $l$, which corresponds to the soft quark $l^2 \sim  m_q^2$.  The
integral over the gauge boson momentum $l_g$ in this case has the same
structure as in the one-loop contribution to the vector Sudakov form factor
{\it i.e.} has the double-logarithmic scaling when  the gauge boson momentum
$l_g$ is soft and the single-logarithmic scaling when  $l_g$ is collinear to
either $p_2$ or $l$.  As for the vector form factor the ultraviolet-divergent
part  of Fig.~\ref{fig::3}(a) is cancelled against the corresponding
ultraviolet-divergent parts of Figs.~\ref{fig::2}(c,d). Then we can decompose
the lower quark propagator as follows
\begin{equation}
S(l+l_g)= S(l+l_g^+)-S(l+l_g)\left(\gamma^+l_g^-+\slashed{l}_g^\perp\right)S(l+l_g^+)\,.
\label{eq::decomp}
\end{equation}
The second term in the above equation can be neglected if $l_g$ is collinear
to $p_2$ or  soft. In the former case $l_g^-\ll l_g^+$ ({\it cf.}
Eq.(\ref{eq::residue})) while $l_g^\perp$ factor makes  the integral over
$l_g^\perp$ nonlogarithmic ({\it cf.} Eq.(\ref{eq::collint3})). For soft
$l_g$ one has   $\left(l_g^\perp\right)^2\sim l_g^-l_g^+$ so the second term
in Eq.(\ref{eq::decomp}) is proportional to $l_g^-$.  Moreover one can
neglect $l_g^2$ in the denominator of the second quark propagator which
becomes $S(p_2+l_g)\sim 1/l_g^-$ and is cancelled by the $l_g^-$ factor so
that the integral over $l_g$ depends on $l^2\sim m_q^2$ and $m_q^2$ only and
does not generate logarithmic corrections.   Note that the above
approximation is not valid for $l_g$  collinear to $l$, which will be
considered separately. In a covariant gauge only $A^-$  light-cone component
of the photon field can be emitted by the eikonal quark line with the
momentum $p_{2}$, while the emission of the $A^+$ and transverse components
is suppressed. Thus the interaction of the virtual photon which is  soft or
collinear to $p_2$  to the quark line  in Fig.~\ref{fig::3}(a) can be
approximated as follows
\begin{equation}
S(l)\gamma^\mu S(l+l_g)\approx S(l)\gamma^-S(l+l_g^+)={1\over l_g^+}
\left(S(l)-S(l+l_g^+)\right)\,.
\label{eq::wardident}
\end{equation}
which is equivalent to the  QED Ward identity.  The  right hand side of
Eq.~(\ref{eq::wardident}) corresponds to the diagram Fig.~\ref{fig::3}(b) where
the crossed circle on the quark propagator represents the replacement $S(l)\to
S(l)-S(l+l_g^+)$ and the $1/l_g^+$ factor is absorbed  into the upper eikonal
quark propagator. By the momentum shift $l\to l-l_g^+$ in the second term of the
above expression the crossed circle can be moved to the upper eikonal photon
line  which becomes ${1\over 2p_1l}-{1\over 2p_1(l+l_g^+)}$,
Fig.~\ref{fig::3}(c). The opposite eikonal line is not sensitive to this shift
since $p_2^-\approx 0$. On the final step we use the  ``inverted Ward identity''
on the upper  eikonal gauge boson line
\begin{equation}
{1\over l_g^+}\left({1\over 2p_1l}-{1\over 2p_1(l+l_g^+)}\right)
={1\over 2p_1l}2{p_1}^-{1\over 2p_1(l+l_g^+)}
\approx {1\over (p_1-l)^2}2p_1^\mu{1\over (p_1-l-l_g)^2}
\label{eq::invident}
\end{equation}
to  transform the diagram Fig.~\ref{fig::3}(c) into Fig.~\ref{fig::3}(d) with an
effective dipole coupling $2e_q p_1^\mu$ to the eikonal {\it gauge boson}, where
 $e_q$ is the {\it quark} charge.  Note that we can replace $2p_1(l+l_g^+)$ by
$-(p_1-l-l_g)^2$  in the gauge boson propagator as long as $l_g\ll Q$ since
$p_1^+\approx 0$.

Then for the soft virtual momentum $l_g$ one can also use the eikonal
approximation for the propagators carrying  the external momenta $p_i$. By
adding the symmetric diagram and the diagram Fig.~\ref{fig::2}(b) we  get a
``ladder'' structure characteristic  to the standard eikonal factorization
for the Sudakov form factor.  This factorization, however,  requires the
summation over all possible insertions of the soft photon vertex along each
eikonal line while in the case under consideration the diagram in
Fig.~\ref{fig::1}(b) with the soft exchange between the  photon  lines is
missing. This diagram can be added to complete the factorization and then
subtracted. After adding the diagram Fig.~\ref{fig::1}(b) the integral over
the soft gauge boson momentum  factors out with respect to the leading order
amplitude. The additional diagram with effective gauge boson interaction
accounts for the variation of the  charge propagating along the eikonal lines
at the point of the soft quark emission and is characteristic  to the leading
mass-suppressed amplitudes \cite{Liu:2017vkm,Liu:2018czl}. At the same time
for the virtual momentum  collinear to $p_2$  the  eikonal approximation can
be used for the propagators carrying  the external momentum $p_1$ while the
integration over  $l_g^-$ is performed by taking the residue  of the
$S(p_2+l_g)$ propagator. Thus the  integration over the  collinear gauge
boson momentum completely factors out in the sum  of the diagrams
Fig.~\ref{fig::2}(b) and  Fig.~\ref{fig::3}(d), as well as in the symmetric
diagrams contribution  when   $l_g$  is collinear to  $p_1$, without any
additional subtraction required for the soft virtual momentum. This can be
easily understood since in contrast to the soft emission the collinear one is
not sensitive to the eikonal charge nonconservation.

In the next-to-leading logarithmic approximation we also need to consider the
case when the virtual quark momentum $l$ is either hard  or collinear and the
corresponding integral is single-logarithmic while the gauge boson momentum
$l_g$ is soft and the corresponding integral has double-logarithmic scaling. For
hard $l$ the integration over $l_g$ trivially factorizes and is
double-logarithmic only for the diagram Fig.~\ref{fig::2}(b). If $l$ is
collinear to $p_2$ the integral over $l_g$ in the  diagram Fig.~\ref{fig::2}(a)
is not double-logarithmic since the soft exchange is between two collinear quark
lines. At the same time for $l$ being collinear to $p_1$ the integral over $l_g$
in this diagram  is double-logarithmic but the above algorithm with the momentum
shift is not applicable since the upper line gauge boson propagator has to be
on-shell. However in this case the diagrams Fig.~\ref{fig::2}(a,b) already give
all possible insertions of the soft gauge boson vertex along the eikonal quark
line collinear to  $p_1$ so the integration over the soft momentum $l_g$ also
factorizes~\cite{Yennie:1961ad}.

The factorized soft and collinear contributions together with the external quark
self-energy diagrams Fig.~\ref{fig::2}(c) add up to the one-loop  contribution
to the universal factor which describes the next-to-leading Sudakov logarithms
for the amplitudes with the  quark-antiquark external on-shell lines
\cite{Korchemsky:1988pn,Kuhn:1999nn,Jantzen:2005az}
\begin{equation}
{Z_q^2}^{NLL}=\exp\left\{-{\alpha_q \over 4\pi}
\left[{2\over\varepsilon}\left({\mu_f^2\over m_q^2}\right)^\varepsilon
\left(1-L\right)+L^2\left(1-{\beta_0\over 3}{\alpha_q \over 4\pi}L\right)\right]
+\gamma_q^{(1)}{\alpha_q \over 2\pi}L\right\}\,,
\label{eq::Zq}
\end{equation}
where the coupling constant is renormalized at the scale  $m_q$, $\beta_0=-4/3$
and  $\gamma_q^{(1)}=3/2$ are the corresponding beta-function and the collinear
quark anomalous dimension, and the ``factorization scale'' $\mu_f$ is the mass
parameter of the dimensional regularization used to deal with  the soft
divergence not regulated by the quark mass. Hence the next-to-leading
logarithmic approximation for the amplitude can be written in the following form
\begin{equation}
{\cal G}^{NLL}= {Z_{q}^2}^{NLL}\left(g(-x)+{\alpha_qL\over 4\pi}
\Delta_q(-x)\right)\,{\cal G}^{(0)}\,.
\label{eq::Gfac}
\end{equation}
In this equation the function $g(-x)$ of the variable $x=-{\alpha_q\over
4\pi}L^2<0$ incorporates the double-logarithmic non-Sudakov contribution of
Fig.~\ref{fig::1}(b) with an arbitrary number of the effective  soft gluon
exchanges \cite{Liu:2018czl}. It is given by the integral
\begin{equation}
g(x)=2\int_0^1 {\rm d}\xi \int_{0}^{1-\xi}{\rm d}\eta e^{2x\eta\xi}\,,
\label{eq::gint}
\end{equation}
where the expression in the exponent is the result of the one-loop integration
over the soft gauge boson virtual momentum in Fig.~\ref{fig::1}(b). The integral
can be solved in terms of generalized hypergeometric function
\begin{equation}
g(x)={}_2F_2\left(1,1;{3/2},2;{x/2}\right)=1+{x\over 6}+{x^2\over 45}+{x^3\over
420}+\ldots\,.
\label{eq::gres}
\end{equation}
and has  the following asymptotic behavior at $x\to+\infty$
\begin{equation}
g(x)\sim \left({2\pi e^{x}\over x^{3}}\right)^{1/2}\!\!.
\label{eq::gasymp}
\end{equation}
The function $\Delta_q(-x)$ in Eq.~(\ref{eq::Gfac}) accounts for the all-order
non-Sudakov next-to-leading  logarithmic corrections. In two loops these
corrections are generated by a part of the collinear contribution which does
not factor into Eq.~(\ref{eq::Zq}) and by the renormalization group logarithms
proportional to beta-function.  Note that  to get the  next-to-leading  two-loop
logarithmic contribution the integration over the virtual quark momentum $l$
should be double-logarithmic and we can use the same approximation as for the
calculation of the one-loop amplitude. Let us consider the collinear
contribution first.  As it has been pointed out the contribution of the virtual
momentum $l_g$ which is collinear to $l$ in the diagram  Fig.~\ref{fig::2}(a)
cannot be factorized into the external quark lines.  Since in the
double-logarithmic approximation the soft quark momentum is close to the mass
shell $l^2\approx m_q^2$ this contribution can be read off the one-loop result
for the  Sudakov massive quark form factor with the external on-shell momenta
$l$ and $p_2$. After adding the symmetric contribution and the self-energy
diagram Fig.~\ref{fig::2}(d) and integrating over $l_g$ we get the factor
\begin{equation}
\gamma_q{\alpha_q\over 4\pi}\left[\ln\left({ (lp_1)\over m_q^2}\right)
+\ln\left({(lp_2)\over m_q^2}\right)\right]
=\gamma^{(1)}_q{\alpha_qL\over 4\pi}\left(2-\eta-\xi\right)\,.
\label{eq::coll2l}
\end{equation}
The  renormalizetion group logarithms from the vacuum polarization of the
off-shell eikonal photon propagators in  Fig.~\ref{fig::2}(e) and the symmetric
diagram read
\begin{equation}
-\beta_0{\alpha_qL\over 4\pi}\left(2-\eta-\xi\right)\,,
\label{eq::b02l}
\end{equation}
where we set the renormalizarion scale of the gauge field operator and
$\alpha_q$ in the leading order amplitude to $m_q$. To get the two-loop cubic
logarithms the expressions  Eqs.~(\ref{eq::coll2l},\ref{eq::b02l}) should be
inserted into the integral Eq.~(\ref{eq::gint}) for $x=0$ corresponding to the
leading order amplitude  which gives
\begin{equation}
\Delta^{\rm 2-loop}_q={4\over 3}\left(\gamma^{(1)}_q-\beta_0\right)\,.
\label{eq::Del2l}
\end{equation}
Since the soft emission from an eikonal line of a given charge factorize and
exponentiate, the higher order next-to-leading logarithmic  corrections
associated with  Eqs.~(\ref{eq::coll2l},\ref{eq::b02l}) can be  obtained by
keeping the exponential factor in the integral Eq.~(\ref{eq::gint}) unexpanded
and the corresponding contribution to $\Delta_q$ reads
\begin{equation}
2\left(\gamma_q^{(1)}-\beta_0\right)\int_0^1
{\rm d}\xi \int_{0}^{1-\xi}{\rm d}\eta \left(2-\eta-\xi\right)e^{2x\eta\xi}
=2\left(\gamma_q^{(1)}-\beta_0\right)\left(g(x)-g_\gamma(x)\right)\,,
\label{eq::Delg}
\end{equation}
where
\begin{equation}
g_\gamma(x)={1\over x}\left[\left({\pi e^x\over 2x}\right)^{1/2}{\rm erf}
(\sqrt{x/2})-1\right]
={1\over 3}\left(1 + {x\over 5} + {x^2\over 35} + {x^3\over 315}+\ldots \right)
\label{eq::gg}
\end{equation}
and ${\rm erf}(x)$ is the error function. At $x\to\infty$ Eq.~(\ref{eq::gg})
has the following asymptotic behavior
\begin{equation}
g_\gamma(x)\sim \left({\pi e^x\over 2x^3}\right)^{1/2}+\ldots\,.
\label{eq::ggasym}
\end{equation}
In  three loops, however, a new source of the  next-to-leading logarithms starts
to contribute, which is related to the renormalization group running of the
coupling constant in the diagram with the effective soft gauge boson exchange,
Fig.~\ref{fig::1}(c). This diagram is needed  to provide the factorization of
the $\beta_0$ term in Eq.~(\ref{eq::Zg}). Note that the diagram includes only the
soft gauge boson exchange. The expression for the two-loop subdiagram with the
vacuum polarization insertion  to the soft gauge boson propagator is given up to
the overall sign by  the result for the  two-loop logarithmic corrections
proportional to $\beta_0$ to the off-shell Sudakov  form factor
Eq.~(\ref{eq::F1})  and reads
\begin{equation}
\beta_0
{\alpha_qL\over 2\pi}x\eta\xi\left({L_\mu\over L}-{\eta+\xi\over 2}\right)\,,
\label{eq::b03l}
\end{equation}
where $L_\mu=\ln(Q^2/\mu^2)$ and $\mu$ is the renormalization scale of
$\alpha_q$ in the double-logarithmic variable $x$. As for
Eqs.~(\ref{eq::coll2l},\ref{eq::b02l}), the corresponding higher order
next-to-leading logarithmic  corrections are obtained by inserting
Eq.~(\ref{eq::b03l}) into the integral Eq.~(\ref{eq::gint}) which gives
\begin{equation}
-\beta_0{\alpha_sL\over 2\pi} \int_0^1{\rm d}\xi
\int_{0}^{1-\xi}{\rm d}\eta (2x\eta\xi)
\left({L_\mu\over L}-{\eta+\xi\over2}\right)e^{2x\eta\xi}
=-\beta_0{\alpha_sL\over 4\pi} g_\beta(x)\,,
\label{eq::Delb0}
\end{equation}
where we introduced  the function
\begin{eqnarray}
g_\beta(x) &=&\left[\left({\pi e^x\over 2x}\right)^{1/2}{\rm erf}
(\sqrt{x/2})-g(x)\right]{L_\mu\over L}
+{3\over 2x}\left[\left(1-{x\over 3}\right)
\left({\pi e^x\over 2x}\right)^{1/2}{\rm erf}(\sqrt{x/2})-1 \right]
\nonumber\\
&=&
\left({x\over 6}+{2\over 45}x^2+{x^3\over 140}+\ldots\right)
{L_\mu\over L}-\left({x\over 15}+{2\over 105}x^2+{x^3\over 315}+\ldots\right)
\label{eq::gb0}
 \end{eqnarray}
with  the following asymptotic behavior at $x\to\infty$
\begin{equation}
g_\beta(x)\sim \left({\pi e^x\over 2x}\right)^{1/2}
\left({L_\mu\over L}-{1\over 2}\right)\,.
\label{eq::gb0asym}
\end{equation}
Note that the leading term  of the expansion in $1/x$, Eq.~(\ref{eq::gb0asym}),
vanishes at the normalization scale $\mu =\sqrt{Qm_q}$.
The complete next-to-leading logarithmic contribution reads
\begin{equation}
\Delta_q(x)=2\left(\gamma^{(1)}_q-\beta_0\right)\left(g(x)-g_\gamma(x)\right)
-\beta_0g_\beta(x)\,.
\label{eq::Deltot}
\end{equation}
We confirm  Eq.~(\ref{eq::Deltot}) in two-loop approximation through the explicit
evaluation of the corresponding  Feynman integrals as functions of  the quark
mass with subsequent  expansion of the  result in $m_q$. A detailed discussion of
such a calculation will be published elsewhere \cite{Anastasiou:2020qzk}.

Finally we should note that according  to Eq.~(\ref{eq::F1}) the collinear
logarithm Eq.~(\ref{eq::coll2l}) and the renormalization group logarithm
Eq.~(\ref{eq::b03l})  exponentiate while the higher order renormalization group
logarithms of the form Eq.~(\ref{eq::b02l})  sum up to the usual running
coupling constant  factors
\begin{equation}
\left[1+\beta_0{\alpha_qL\over 4\pi}\left(1-{\eta}\right)\right]^{-1}
\left[1+\beta_0{\alpha_qL\over 4\pi}\left(1-{\xi}\right)\right]^{-1}\,.
\end{equation}
Thus we can rewrite Eq.~(\ref{eq::Gfac}) in a more orthodox  form
\begin{equation}
{\cal G}^{NLL}= 2\int_0^1 {\rm d}\xi \int_{0}^{1-\xi}{\rm d}\eta\,
{e^{\left\{-2x\eta\xi\left[1-\beta_0
{\alpha_qL\over 4\pi}\left({L_\mu\over L}-{\eta+\xi\over 2}\right)\right]
+\gamma^{(1)}_q{\alpha_qL\over 4\pi}\left(2-\eta-\xi\right)
\right\}}
\over \left[1+\beta_0{\alpha_qL\over 4\pi}\left(1-{\eta}\right)\right]
\left[1+\beta_0{\alpha_qL\over 4\pi}\left(1-{\xi}\right)\right]}
\,{Z_{q}^2}^{NLL}{\cal G}^{(0)}\,,
\label{eq::Gnll}
\end{equation}
which would naturally appear within an effective field theory analysis and
includes also the terms $\alpha_q^nL^{m}$  with $m<2n-1$.   We however prefer
to work with the strict logarithmic expansion,
Eqs.~(\ref{eq::Gfac},\ref{eq::Deltot}). These equations reflect  the general
structure of the non-Sudakov next-to-leading logarithmic corrections and will
be generalized to the Higgs boson production in the next section.

\section{$gg\to H$ amplitude mediated by a light quark}
\label{sec::3}
\begin{figure}
\begin{center}
\begin{tabular}{cccc}
\hspace*{03mm}\includegraphics[width=1.8cm]{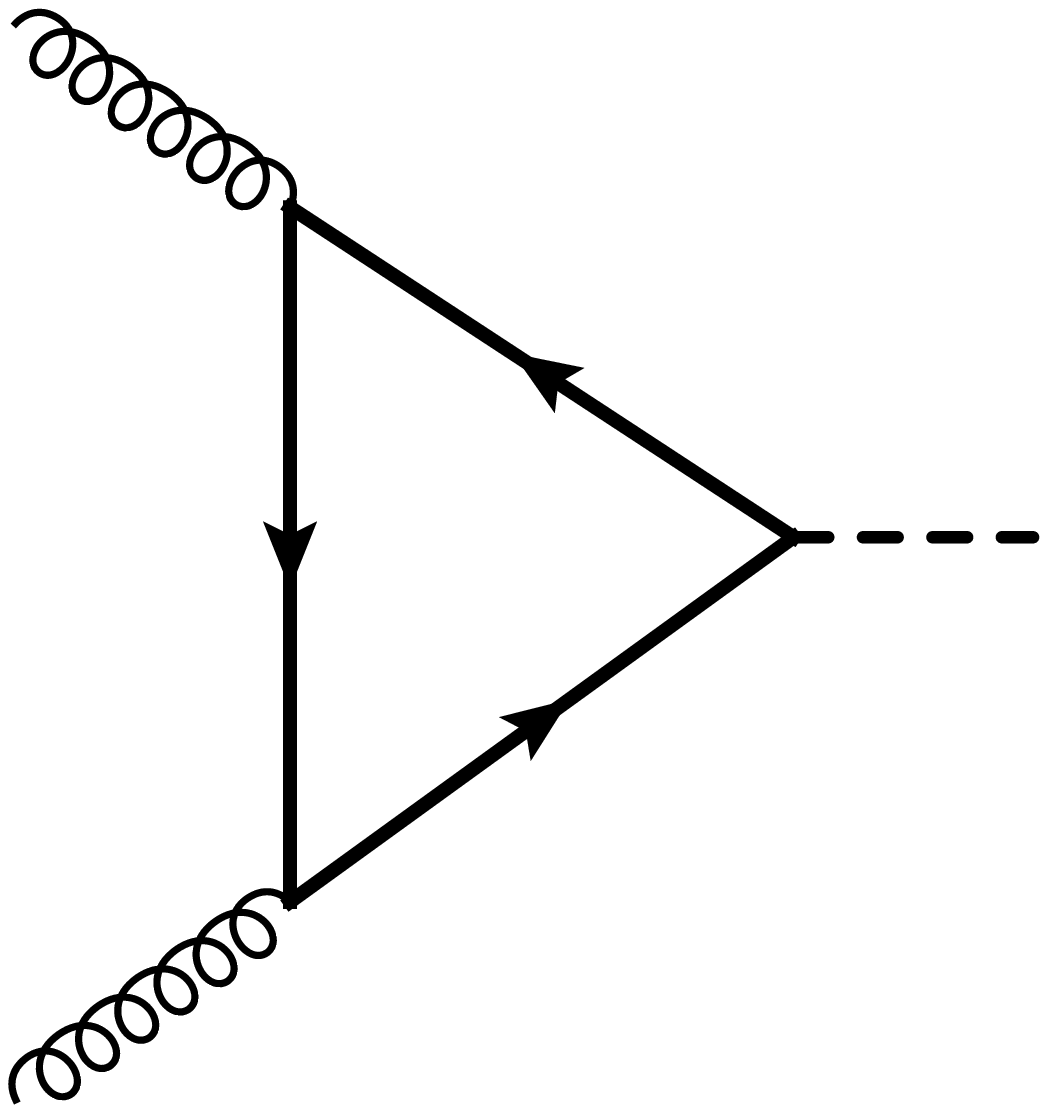}&
\hspace*{03mm}\includegraphics[width=1.8cm]{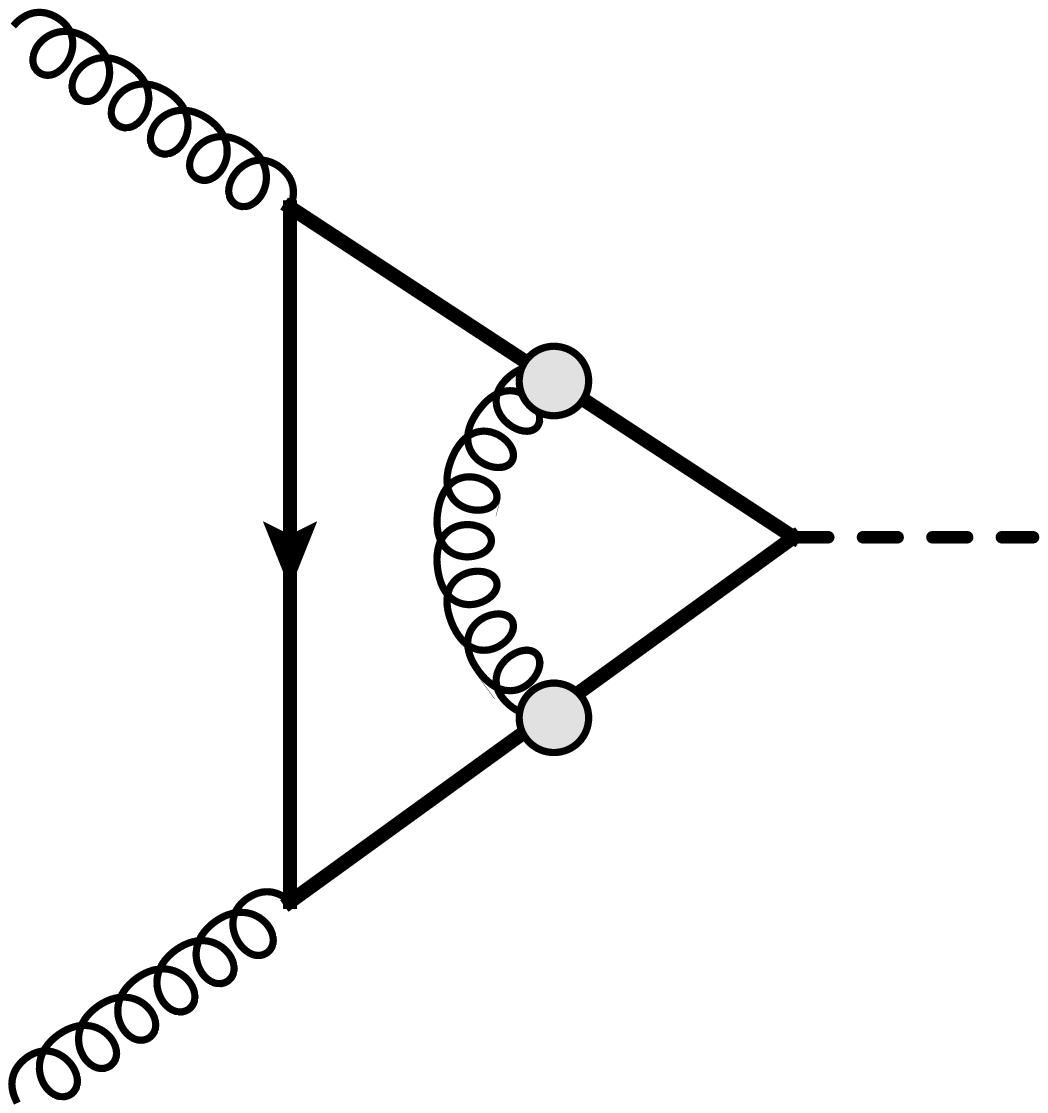}&
\hspace*{03mm}\includegraphics[width=1.8cm]{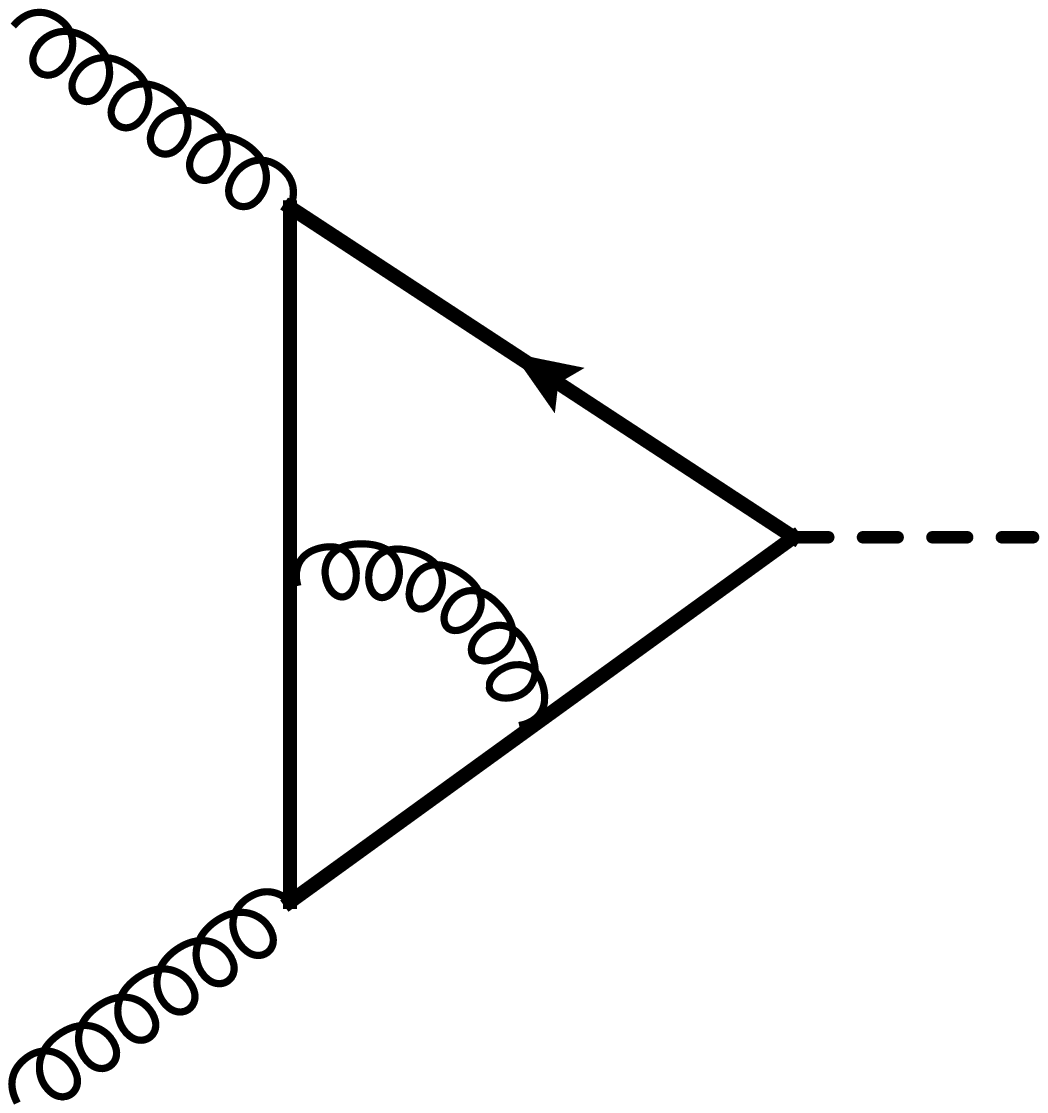}&
\hspace*{03mm}\includegraphics[width=1.8cm]{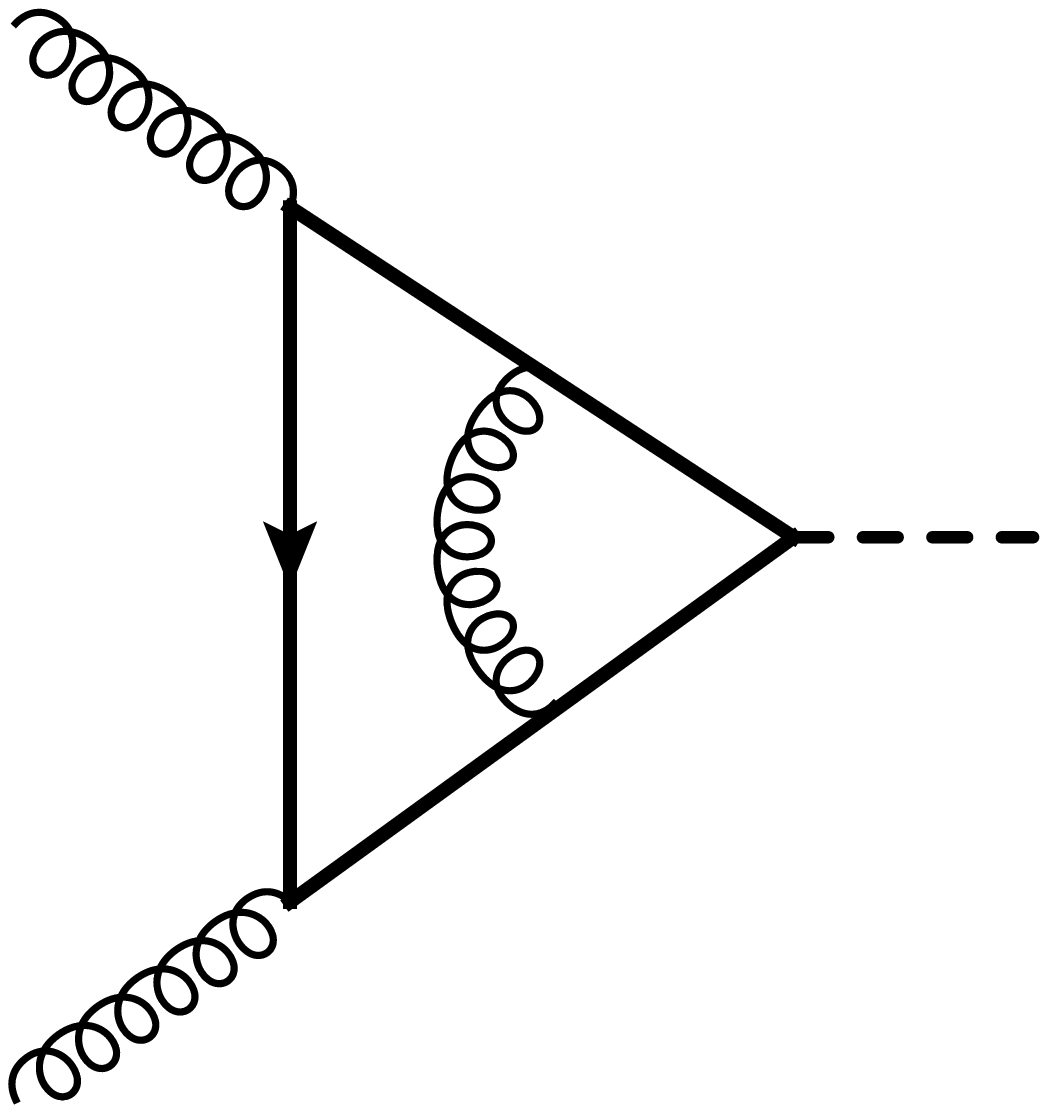}\\
(a)&\hspace*{03mm}(b)&\hspace*{03mm}(c)&\hspace*{03mm}(d)\\
\end{tabular}
\end{center}
\caption{\label{fig::4} The Feynman diagram for (a) the leading order one-loop
Higgs boson production in gluon fusion,  (b) the effective soft gauge boson
exchange similar to Fig.~\ref{fig::1}(b), (c) gluon and (d) Higgs boson vertex
corrections.}
\end{figure}

The leading order amplitude of the gluon fusion into the Higgs boson is given by
the one-loop diagram in Fig.~\ref{fig::4}(a). The dominant contribution to the
process  comes from the top quark loop and in the formal limit of the large top
quark mass $m_t\gg m_H$ is proportional  to the square of the Higgs boson mass
$m_H$. By contrast for a light quark  with $m_q\ll m_H$ running inside the loop
the amplitude is proportional to $m_q^2$.  Indeed, the  Higgs boson coupling to
the quark is proportional to  $m_q$.  Then the scalar interaction of the Higgs
boson results in a helicity flip at the interaction vertex and helicity
conservation requires the amplitude to vanish in the limit $m_q\to 0$ even if
the Higgs coupling to the light quark is kept fixed. By using the explicit
one-loop result the light quark mediated amplitude can be written in such a way
that its power suppression and the logarithmic enhancement  is manifest
\begin{equation}
{{\cal M}^q}^{(0)}_{gg\to H}=-{3\over 2}{m_q^2\over m_H^2}L^2 \,
{{\cal M}^{t}}^{(0)}_{gg\to H}\,,
\label{eq::MH0}
\end{equation}
where $L=\ln(-s/m_q^2)$, $s\approx m_H^2$ is the
total energy of colliding gluons, and the result is given in terms of the heavy
top quark  mediated  amplitude ${{\cal M}^{t}}^{(0)}_{gg\to H}$, which
corresponds to a local gluon-gluon-Higgs interaction vertex and has one
independent helicity component.

Though apparently completely different, the $gg\to H$ amplitude shares several
crucial  properties with  the quark scattering amplitude  considered in the
previous section: it is suppressed by the  leading power of the quark mass due
to the helicity flip,    the double-logarithmic contribution is induced by the
soft quark exchange  and the (color) charge is not conserved along the eikonal
lines.  Moreover, since the eikonal  (Wilson) lines are characterized by the
momentum and color charge but not the spin, the factorization structure of soft
and collinear logarithmic corrections described  in the Sect.~\ref{sec::2.2}
directly applies to the case under consideration.  In particular the non-Sudakov
double-logarithmic corrections are determined by the diagram
Fig.~\ref{fig::4}(b) with the effective soft gluon exchange and are described by
the same function $g(x)$ with $x=(C_A-C_F){\alpha_s\over 4\pi}L^2>0$, where the
color factor accounts for the eikonal color charge variation and the opposite
sign of the argument is dictated by the direction of the color
flow.\footnote{The relation between the double-logarithmic asymptotic behavior
of different amplitudes and gauge theories is discussed in detail in
Refs.~\cite{Liu:2017vkm,Liu:2018czl}} Thus the next-to-leading factorization
formula for the amplitude takes the following form ({\it cf.}
Eq.~(\ref{eq::Gfac}))
\begin{equation}
{{\cal M}^q}^{NLL}_{gg\to H}= {Z^2_{g}}^{NLL} \left(g(x)+
{\alpha_sL\over 4\pi}\Delta_g(x)\right)
\left({\alpha_s(m_H)\over \alpha_s(m_q)}
\right)^{\gamma_m^{(1)}/\beta_0}\,{{\cal M}^q}^{(0)}_{gg\to H}\,,
\label{eq::Mfac}
\end{equation}
where the Sudakov factor for  a gluon scattering is (see {\it e.g.}
\cite{Catani:1998bh})
\begin{equation}
 {Z^2_{g}}^{NLL}=\exp\left[-{\alpha_s\over 4\pi}\left({2C_A
 \over\varepsilon^2}+{\beta_0\over \varepsilon}\right)
 \left({s\over \mu^2}\right)^{-\varepsilon}
 +{\cal O}(\alpha_s^2)\right]\,,
\label{eq::Zg}
 \end{equation}
and we renormalize the strong coupling constant in the leading order amplitude
at the infrared factorization scale $\mu_f=\mu$. The extra renormalization group
factor appears in  Eq.~(\ref{eq::Mfac}) since  the leading order amplitude is
defined in terms of the quark pole mass while the physical renormalization
scale of the quark Yukawa coupling is $m_H$.

As for the quark scattering amplitude, the next-to-leading
logarithmic term $\Delta_g(x)$  gets contributions from the collinear
gluon exchange which does not factorize into  Eq.~(\ref{eq::Zg}) and
from the  renormalization group running of the coupling constant in
the diagram Fig.~\ref{fig::4}(b) with the effective soft gluon
exchange. The latter is identical to Eq.~(\ref{eq::Delb0}) up to the
definition of $x$ variable and the value of beta-function. The
nonfactorizable  collinear contribution  needs additional analysis
though. The diagram which can potentially develop such a contribution
can be classified as the corrections to the gluon  and Higgs boson
vertices  with the typical examples given in Fig.~\ref{fig::4}(c,d).
In these diagrams the integral over the soft quark momentum  $l$
should be evaluated in the double-logarithmic approximation which
effectively put the soft quark on its mass shell $l^2\approx m_q^2$.
The correction to the gluon vertices, Fig.~\ref{fig::4}(c),  with one
of the quark lines off-shell by the amount $(lp_i)$ and the
corresponding quark self-energy result in a single collinear
logarithm
\begin{equation}
\gamma_q{\alpha_q\over 4\pi}\left[2\ln\left({ (p_1p_2)\over (p_1l)}\right)
+\ln\left({ m_q^2\over (p_1l)}\right)
+(p_1\leftrightarrow p_2)\right]
=\gamma^{(1)}_q{\alpha_qL\over 4\pi}\left(3\eta+3\xi-2\right)\,.
\label{eq::coll2lg}
\end{equation}
Similar contribution associated with the  Higgs boson vertex
corrections, Fig.~\ref{fig::4}(d) can be read off the
expansion of the Sudakov scalar form factor  Eq.~(\ref{eq::FS})
\begin{equation}
\gamma_q{\alpha_q\over 4\pi}\left[\ln\left({ (p_1p_2)\over (p_1l)}\right)
+\ln\left({ (p_1p_2)\over (p_2l)}\right)\right]
=\gamma^{(1)}_q{\alpha_qL\over 4\pi}\left(\eta+\xi\right)\,.
\label{eq::coll2lh}
\end{equation}
Note that the renormalization group running of the scalar coupling has alredy
been taken into account in the factorization formula   Eq.~(\ref{eq::Mfac}). To
get the corresponding all-order corrections the sum of
Eqs.~(\ref{eq::coll2lg},\ref{eq::coll2lh}) should be inserted
into the integral representation of the function $g(z)$ which gives the
following contribution\footnote{The nonabelian gauge interaction
does  not affect the factorization and exponentiation of the double logarithmc
contribution \cite{Frenkel:1984pz}.} to $\Delta_g$
\begin{equation}
2\gamma_q^{(1)}\int_0^1
{\rm d}\xi \int_{0}^{1-\xi}{\rm d}\eta \left(4\eta+4\xi-2\right)e^{2x\eta\xi}
=2\gamma_q^{(1)}\left(4g_\gamma(x)-g(x)\right)\,.
\label{eq::Delgh}
\end{equation}
Thus the total result for the next-to-leading logarithmic contribution to the
amplitude reads
\begin{equation}
\Delta_g(x)=2\gamma^{(1)}_q \left(4g_\gamma(x)-g(x)\right)-\beta_0g_\beta(x)
\label{eq::Deltoth}
\end{equation}
with the following perturbative expansion
\begin{eqnarray}
\Delta_g(x)&=&C_F\left(1+{3\over 10}x+{x^2\over 21}+\ldots\right)
-\beta_0\left[\left({x\over 6}+{2\over 45}x^2+\ldots\right){L_\mu\over L}\right.
\nonumber\\
&-&\left.\left({x\over 15}+{2\over 105}x^2+\ldots\right)\right]\,,
\label{eq::Delseriesh}
\end{eqnarray}
where $L_\mu=\ln(-s/\mu^2)$ and the functions $g_\gamma(x)$ and
$g_\beta(x)$ are defined by Eq.~(\ref{eq::gg}) and
Eq.~(\ref{eq::gb0}), respectively.  In the above equation the
three-loop contribution proportional to the beta-function vanishes
when the strong coupling constant in the double-logarithmic variable
$x$  is renormalized at the scale $\mu
=m_H\left({m_q/m_H}\right)^{2/5}$, which can be considered as the
``optimal'' renormalization scale in this case.

\section{Higgs boson threshold production}
\label{sec::4}
The perturbative analysis  of the total  cross section of
the Higgs boson production requires the inclusion  of the
real  emission contribution, which is not yet available for
the light quark loop mediated process with arbitrary energy
of the emitted partons beyond the next-to-leading order
approximation. However, near the production threshold where
$z=m_H^2/s\to1$ only the soft real emission contributes to
the inclusive cross section. For $1-z\ll m_b/m_H$ the energy
of an emitted gluon  $E_g$ is much less than $m_b$. Up to
the corrections suppressed by $E_g/m_b\ll1$ such  emission
does not resolve the bottom quark  loop and has the same
structure as  in the top quark loop mediated process, where
it is known through the next-to-next-to-next-to-leading
order. Thus   we can apply our result for the analysis  of
the higher order corrections to the Higgs boson threshold
production.

\subsection{Partonic cross section near the threshold}
\label{sec::4.1}
The expansion  of  Eqs.~(\ref{eq::Mfac},\ref{eq::Deltoth})   gives
\begin{eqnarray}
{{\cal M}^b}_{gg\to H}&=& C_b
Z^2_{g}\left({\alpha_s(m_H)\over
\alpha_s(m_b)}\right)^{\gamma_m^{(1)}/\beta_0}\,
{{\cal M}^b}^{(0)}_{gg\to H}\,,
\label{eq::3loopM}
\end{eqnarray}
where $C_b=1+\sum_{n=1}^\infty c_n$ and  up to  four  loops in the
next-to-leading logarithmic approximation  we get\footnote{Since the imaginary
part of the amplitude does not contribute to the cross section in the
next-to-leading logarithmic approximation we neglect the  imaginary part of the
logarithms and define  $L=\ln(m_H^2/m_b^2)$, $L_\mu=\ln(m_H^2/\mu^2)$ in the
rest of the paper}
\begin{eqnarray}
c_1&=&{x\over 6}+C_F{\alpha_s L\over 4\pi}\,,\nonumber\\
c_2&=&{x^2\over 45}+{x\over 5}{\alpha_s L\over 4\pi}
\left[{3\over 2}C_F-\beta_0\left({5\over 6}{L_\mu\over L}
-{1\over 3}\right)\right]\,,\nonumber\\
c_3&=&{x^3\over 420}+{x^2\over 5}{\alpha_s L\over 4\pi}
\left[{5\over 21}C_F-\beta_0\left({2\over 9}{L_\mu\over L}
-{2\over 21}\right)\right]\,.
\label{eq::Cb}
\end{eqnarray}
The  coefficient $c_1$ agrees with the small-mass expansion of the exact
two-loop result  (see {\it e.g.} \cite{Aglietti:2006tp}). The higher-order
leading  logarithmic terms   have been  obtained  in
Refs.~\cite{Liu:2017vkm,Liu:2018czl} while the next-to-leading logarithmic
terms are new. The $n_l$ part of the  $\beta_0$ term in the coefficient $c_2$
agrees with the small-mass expansion of  the analytical three-loop result
\cite{Harlander:2019ioe}.

The three-loop leading logarithmic term  has been recently
confirmed through numerical calculation
\cite{Czakon:2020vql}. Eq.~(\ref{eq::Cb}) corresponds to
the coefficient $(C_A-C_F)^2/5760$ of the $L_s^6/z$  term
in Eq.~(C.1) of \cite{Czakon:2020vql}, which agrees with
the numerical value $0.0004822530864$ given in this paper.
The three-loop next-to-leading logarithmic term however
depends on the ultraviolet renormalization and infrared
subtraction scheme. We keep the bottom quark as an active
flavor since it contributes to the running of the strong
coupling constant  in the relevant scale interval $m_b<\mu
<m_H$. At the same time in Ref.~\cite{Czakon:2020vql}  a
massive quark is treated separately and is decoupled even
for $m_q\ll m_H$. This  changes  the running of the
effective coupling constant as well as  the form of the
infrared divergent factor Eq.~(\ref{eq::Zg}) and makes the
comparison of the results not quite straightforward beyond
the leading logarithmic approximation. As far as we can
conclude the  conversion of Eq.~(\ref{eq::Cb}) to the
scheme used in \cite{Czakon:2020vql} gives the coefficient
$(C_A-C_F)(11C_A/9-C_F-10T_F/3)/640$ of the $L_s^5/z$  term
in Eq.~(C.1), which is in agreement with the numerical
value  $0.001736111111$ obtained in \cite{Czakon:2020vql}.
The mass dependence of the three-loop amplitude has been
also studied by  numerical method based on conformal
mapping and Pad\'e approximants in
Ref.~\cite{Davies:2019nhm} but the accuracy of this method
in the small-mass limit is not sufficient for a comparison
with our result.

The dominant contribution to the cross section is due to
the  interference of ${{\cal M}^b}_{gg\to H}$  with the top
quark loop mediated amplitude. We can define the
nonlogarithmic part of the gluon Sudakov factor  $Z^2_{g}$
in such a way that it coincides with the gluon form factor
which accounts for the virtual corrections to the $gg\to H$
amplitude  in the heavy top quark effective theory. After
combining it  with the soft real emission we get the known
effective theory expression for the radiative corrections
to the top quark loop mediated threshold cross section.
Then we can write the above top-bottom interference
contribution  in the factorized form
\begin{eqnarray}
\delta\sigma_{gg\to H+X}(s)&=& -3{m_b^2\over m_H^2}\left({\alpha_s(m_H)\over
\alpha_s(m_b)}\right)^{\gamma_m^{(1)}/\beta_0}L^2\,
C_b\,C_t\,\sigma^{\rm eff}_{gg\to H+X}\,.
\label{eq::delsigma}
\end{eqnarray}
In Eq.~(\ref{eq::delsigma}) $m_b$ is the bottom quark pole mass,  the heavy
top quark effective theory Wilson coefficient for $N_c=3$ reads
\cite{Chetyrkin:1997un,Schroder:2005hy,Chetyrkin:2005ia}
\begin{equation}
C_t=1+{11\over 4}{\alpha_s\over\pi}+\left({\alpha_s\over\pi}\right)^2
\left[{2777\over 288}-{19\over 16}\ln\left({m_t^2\over \mu^2}\right)-
n_l\left({67\over 96}+{1\over 3}\ln\left({m_t^2\over \mu^2}\right)\right)\right]
+\ldots\,,
\label{eq::Ct}
\end{equation}
where $\alpha_s$ is the $\overline{\rm MS}$ coupling constant renormalized at
the scale  $\mu$, and the effective theory threshold cross section has the
following perturbative expansion
\begin{eqnarray}
\sigma^{\rm eff}_{gg\to H+X} = {\pi\over v^2(N_c^2-1)^2}
\left({\alpha_s \over 3\pi }\right)^2\sum_{n=0}^\infty
\left({\alpha_s\over \pi}\right)^n\sigma_n\,,
\label{eq::sigmaeff}
\end{eqnarray}
where $v\approx 246$~GeV  is the vacuum expectation value of the Higgs field.
The coefficients of the expansion Eq.~(\ref{eq::sigmaeff}) are the same as for
the top quark loop mediated threshold cross section.  The corresponding
expressions in terms of delta- and plus-distributions up to $n=2$ for $N_c=3$
and the factorization scale $\mu_f=\mu$ read
\cite{Catani:2001ic,Harlander:2001is}
\begin{eqnarray}
\sigma_0 &=& \delta(1-z)\,, \nonumber \\
\sigma_1 &=& 6 \zeta_2\,  \delta(1-z)
+6 L_\mu \DplusZero + 12 \DplusOne  ,
\nonumber\\
\sigma_2&=&  \delta(1-z) \Bigg\{\frac{837}{16}
+\frac{67}{2}\zeta _2
-\frac{165 }{4}\zeta _3-\frac{9}{8}\zeta_4
+\left(-\frac{27}{2}-\frac{33 }{2}\zeta _2
+\frac{171}{2}\zeta _3\right)L_\mu
-18\zeta _2L^2_\mu
\nonumber\\
&+&n_l\left[-\frac{247}{36}-\frac{5 }{3}\zeta _2
+\frac{5}{6}\zeta _3
+\left(\frac{11}{6}+\zeta _2\right)L_\mu \right]
\Bigg\}+\DplusZero \Bigg[-\frac{101}{3}+
33 \zeta _2+\frac{351}{2}\zeta _3
\nonumber\\
&+&\left(\frac{67}{2}-45\zeta _2\right)L_\mu
-\frac{33}{4}L^2_\mu
+n_l\left(\frac{14}{9}-2\zeta _2-
\frac{5}{3}L_\mu
+\frac{1}{2}L^2_\mu\right)\Bigg]
\nonumber\\
&+&\DplusOne \Bigg[67-90 \zeta _2-33L_\mu
+36L_\mu^2 +n_l
\left(-\frac{10 }{3}+2L_\mu\right)\Bigg]
\nonumber\\
&+&\Dplus{2} \left(-33+108L_\mu+2n_l
\right)
+ 72\Dplus{3},
\label{eq::sigman}
\end{eqnarray}
where $\zeta_n=\zeta(n)$ is a value of  Riemann zeta-function. Though
Eq.~(\ref{eq::sigman}) includes terms beyond the next-to-leading logarithmic
accuracy we prefer to keep them since the logarithmic expansions for $c_n$
and $\sigma_n$ are of  quite different nature. Note that in the effective
theory cross section Eq.~(\ref{eq::sigmaeff}) we can take the massless bottom
quark limit and in Eq.~(\ref{eq::sigman}) the number of active flavors is
$n_l=5$. By re-expanding Eq.~(\ref{eq::delsigma}) in $\alpha_s$ we get the
next-to-leading logarithmic result for the leading order (LO),
next-to-leading oder (NLO), and the next-to-next-to-leading order (NNLO)
contributions
\begin{eqnarray}
\delta\sigma^{\rm NLL,\,LO}_{gg\to H+X}&=&N\sigma_0\,,
\nonumber\\
\delta\sigma^{\rm NLL,\,NLO}_{gg\to H+X}&=&
N\left[\left(c_1-3C_F{\alpha_s(\nu)L_\nu\over 4\pi}\right)\sigma_0
+{\alpha_s\over \pi}\sigma_1\right]\,,
\label{eq::sigmanll}\\
\delta\sigma^{\rm NLL,\,NNLO}_{gg\to H+X}&=&
N\Bigg\{\left[c_2-C_F  \frac{\alpha_s(\nu)L_\nu}{4 \pi}{x\over 2}\right]\sigma_0
+{\alpha_s\over \pi}\left[c_1-3C_F{\alpha_s(\nu)L_\nu\over 4\pi}\right]\sigma_1
+\left({\alpha_s\over \pi}\right)^2\sigma_2\Bigg\}\,,
\nonumber
\end{eqnarray}
where $L_\nu=\ln(m_H^2/\nu^2)$, the normalization factor is
\begin{eqnarray}
N&=&-{C_t\over 3\pi}{y_b(\nu)\over y_b(m_b)}
\left({\alpha_s L m_b\over  (N_c^2-1)v m_H}\right)^2\,,
\label{eq::N}
\end{eqnarray}
$y_b(\nu)$ is the bottom quark Yukawa coupling renormalized at the scale $\nu$,
and we use  the expansion
\begin{equation}
\left({\alpha_s(m_H)\over
\alpha_s(m_b)}\right)^{\gamma_m^{(1)}/\beta_0}={y_b(\nu)\over y_b(m_b)}
\left[1-3C_F{\alpha_s(\nu)L_\nu\over 4\pi}
+\ldots\right]\,.
\label{eq::yrgseries}
 \end{equation}
The expression for the next-to-next-to-next-to-leading order  (N$^3$LO)
contribution is rather cumbersome and   can be obtained  from
Eq.~(\ref{eq::delsigma}) by using our result for $c_3$ and  the N$^3$LO
threshold  cross-section from \cite{Anastasiou:2014vaa}.

\subsection{Hadronic cross section in  the threshold approximation}
\label{sec::4.2}

With the partonic result from  the previous section we can
estimate  the bottom quark loop contribution to the
hadronic cross section given by
\begin{equation}
\delta\sigma_{pp\to H+X}= \int dx_1\, dx_2\, f_g(x_1)\,f_g(x_2)\,
\delta\sigma_{gg\to H+X}(x_1\,x_2\,s)\,,
\label{eq::sigmahad}
\end{equation}
where  $f_g(x_i)$  is the gluon  distribution function and
$s$ denotes the square of the partonic center-of-mass
energy. In addition to the expressions of the last subsection,
as numerical input, we use the hadronic cross section
coefficients for top quark contributions in the infinite
mass limit evaluated in the threshold approximation through
N$^3$LO as obtained by  {\tt ihixs2}~\cite{Dulat:2018rbf}.

\begin{table}[t]
\begin{center}
\begin{tabular}{|c|c|c|c|c|} \hline
                                   & LO & NLO & NNLO & N$^3$LO       \\ \hline
$\delta\sigma^{\rm LL}_{pp\to H+X}$    & -1.420  & -1.640 & -1.667 &-1.670\\ \hline
$\delta\sigma^{\rm NLL}_{pp\to H+X}$   & -1.420 &  -2.048 &
                                               -2.183 & -2.204      \\ \hline
$\delta\sigma_{pp\to H+X}$   & -1.023 & -2.000 &  & \\ \hline
\end{tabular}
\end{center}
\caption{\label{tab::xhadronic_ihixs}
The bottom quark loop corrections in picobarns to the Higgs
boson production cross section of different orders in
$\alpha_s$ given in  the leading logarithmic approximation
(LL), the  next-to-leading logarithmic approximation (NLL)
and with full dependence on $m_b$. All the results are
obtained  with the threshold partonic cross section at
center of mass energy of 13 TeV and
renormalization/factorization scales set equal to {\em
half} the Higgs boson mass. Following the conventions of
Ref.~\cite{Anastasiou:2016cez}, we use the values of the
top and bottom quark Yukawa couplings in the
$\overline{{\rm MS}}$-scheme.  Our input values at
$\mu=m_H/2$ are ${\overline m}_b(\mu)=2.961\, {\rm GeV}$
and $\alpha_s(m_H/2)=0.1252$. The top quark mass is set to
a very large value.}
\end{table}

The numerical results for the top-bottom interference
contribution to the cross section in  different orders in
$\alpha_s$ are presented in
Table~\ref{tab::xhadronic_ihixs} for the following values
of input parameters: $\alpha_s(M_Z)=0.118$, $n_l=5$,
$\nu=\mu_f=\mu=m_H/2$, ${\overline m}_b({\overline
m}_b)=4.18$~GeV, $m_H=125$~GeV. The above choice of  $\mu$
ensures a good convergence of the series
Eq.~(\ref{eq::sigmaeff}) for $\sigma^{\rm eff}_{gg\to H+X}$
~\cite{Anastasiou:2016cez}. At the same time $\sigma^{\rm
eff}_{gg\to H+X}$ and $C_b$ in Eq.~(\ref{eq::delsigma}) are
separately renormalization group invariant and in general
one can use a different value of $\mu$ in the series for
$C_b$,  Eq.~(\ref{eq::Cb}). However, the corresponding
optimal value $\mu =m_H\left({m_b/ m_H}\right)^{2/5}\approx
m_H/3$  is quite close to the one for $\sigma^{\rm
eff}_{gg\to H+X}$ and therefore we use the same
renormalization scale in both series. Note that in the NLL
approximation there is no difference between the pole mass
$m_b$ and the $\overline{{\rm MS}}$ mass ${\overline
m}_b({\overline m}_b)$ and we use the latter as the bottom
quark mass parameter for its better perturbative
properties.

\begin{figure}[t]
\begin{center}
\includegraphics[width=0.8\textwidth]{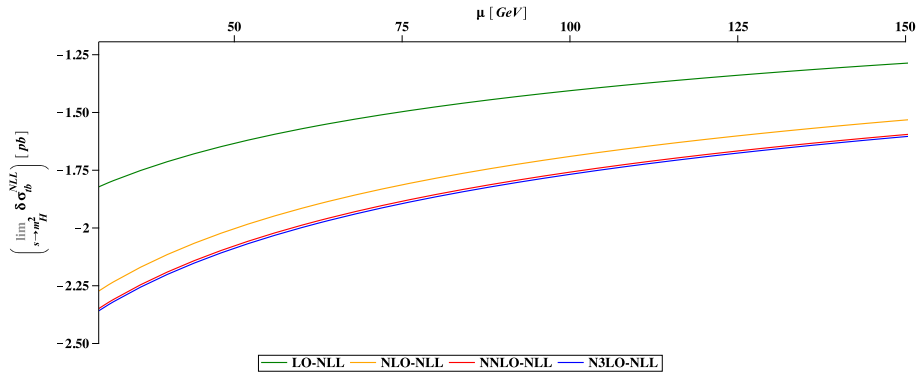}
\end{center}
\caption{\label{fig:sigmaNLL}
The scale dependence of the  top-bottom interference
contribution to the threshold cross section in the NLL
approximation. The factorization scale and the
renormalization scales of the bottom quark Yukawa coupling
and the strong coupling constant are set equal
$\mu_f=\nu=\mu$. The bottom quark mass, other than in the
Yukawa coupling,  is set equal to ${\overline
m}_b({\overline m}_b)$. This is a natural choice and
differs from the one in Table~\ref{tab::xhadronic_ihixs}
where for purposes of comparison with prior literature we
used ${\overline m}_b(\mu)$ universally.
}
\end{figure}

In Table~\ref{tab::xhadronic_ihixs} we also present the
result obtained with the threshold partonic cross section
retaining full dependence on  $m_b$, which is available up
to NLO, and the leading logarithmic result obtained with
the same $\sigma_n$ coefficients but all the subleading
terms in Eqs.~(\ref{eq::Cb},\ref{eq::yrgseries})  being
neglected. As we see, both perturbative and logarithmic
expansions have a reasonable convergence. In LO and NLO,
where the full mass dependence is known, we find that the
NLL cross section is within $42\%$ and  $3\%$ of the exact
result, respectively. The inclusion of the NLL terms is
crucial for reducing the scale dependence  as it determines
the scales of the bottom quark mass, Yukawa coupling and
the strong coupling constant in the LL result. For example,
the renormalization group invariant product of the Yukawa
factor  Eq.~(\ref{eq::yrgseries}) and the coefficient $C_b$
in NNLO is decreased by about $17\%$ with the scale
variation from $m_H/3$ to $2m_H$ in the LL approximation
and only by about $5\%$ in the NLL one. The scale
dependence of the different orders of perturbative
expansion for the threshold cross section in the NLL
approximation is shown in Fig.~\ref{fig:sigmaNLL}.

We should emphasize that the results in
Table~\ref{tab::xhadronic_ihixs} and
Fig.~\ref{fig:sigmaNLL} are obtained in the threshold
limit.  As discussed, for example,  in
Ref.~\cite{Anastasiou:2014lda}, threshold corrections are
not uniquely defined for the hadronic cross section
integral. Diverse definitions lead to important numerical
differences in the estimate of the cross section. In this
paper  we have adopted the simplest choice of a flux for
the partonic cross section which is given by
Eq.~(\ref{eq::sigmahad}).   For top quark contribution
only, with the same flux choice,  the threshold N$^3$LO
cross section in the infinite top quark mass limit
constitutes $\sim 65\%$ of the full cross section. While
the threshold contribution may not be adequate for precise
estimate of the cross section, it does constitute a
physical quantity (in contrast to infrared divergent
amplitudes) and can therefore be used to detect whether the
large logarithms pose any challenges for the convergence of
the perturbative expansion. From the above numerical
results we conclude that while  in  NLO the subleading
logarithms are sizable, beyond  NLO  the logarithmically
enhanced corrections are modest and  under control.

\begin{figure}[t]
\begin{center}
\includegraphics[width=0.8\textwidth]{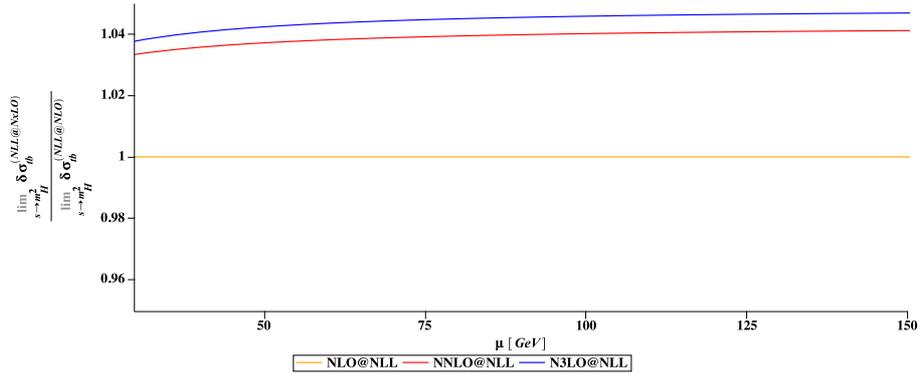}
\end{center}
\caption{\label{fig:RatioNLL}
The scale dependence of the ratio of the top-bottom
interference contribution to the threshold cross section to
the  NLO result through N$^3$LO   in the NLL approximation.
The factorization scale and the renormalization scales of
the bottom quark Yukawa coupling and the strong coupling
constant are set equal $\mu_f=\nu=\mu$. The bottom quark
mass, other than in the Yukawa coupling,  is  set equal to
${\overline m}_b({\overline m}_b)$. }
\end{figure}

In Fig.~\ref{fig:RatioNLL} we plot the ratios of the NNLO
and N$^3$LO top-bottom interference contribution to the
threshold cross section to the  NLO result in the NLL
approximation. For a wide range of scales the K-factors are
in the interval from 1.03 to 1.04.   To get an estimate of
the total NNLO correction to the bottom quark contribution
we can apply the corresponding NLL K-factor to the  NLO
result with  full dependence on  $m_b$ which gives
\begin{equation}
\left({\delta\sigma^{\rm NLL,\,NNLO}_{gg\to H+X}
\over \delta\sigma^{\rm NLL,\,NLO}_{gg\to H+X}}-1
\right)\delta\sigma^{\rm NLO}_{gg\to H+X}\approx -0.13~pb\,,
\label{eq::Knnlo}
\end{equation}
where we use the numerical values from
Table~\ref{tab::xhadronic_ihixs}. Similar procedure
gives the N$^3$LO correction of $-0.02~pb$.

The accuracy of the NLL approximation in LO and NLO is
rather good.  For a rough estimate of its accuracy in NNLO
we can use the  sum of the (highly scheme dependent!)
subleading  $L_s^n/z$  three-loop terms with $n=0,\ldots,4$
in Eq.~(C.1) of \cite{Czakon:2020vql}. By applying the
corresponding K-factor to the LO result
$\delta\sigma_{pp\to H+X}$ in
Table~\ref{tab::xhadronic_ihixs} we get a correction of
$0.049~pb$, which constitutes approximately $-40\%$ of the
NLL correction Eq.~(\ref{eq::Knnlo}), very much like for
the LO terms. This tempts  us to assign in general a $40\%$
uncertainty to the NLL approximation. However, the analysis
of the electroweak Sudakov logarithms
\cite{Kuhn:2001hz,Jantzen:2005az} quite similar to the mass
logarithms discussed in this paper suggests that in NNLO
the next-to-next-to-leading logarithms can be numerically
equal to the LL and the NLL terms. This gives us a
conservative estimate of $100\%$ uncertainty of the result
Eq.~(\ref{eq::Knnlo}). Assuming also $15\%$ uncertainty due
to the  N$^3$LO and higher order corrections together with
the $50\%$ uncertainty of the threshold approximation
discussed above and adding up the errors linearly we obtain
a rough estimate of the bottom quark mediated contribution
to the total cross section of Higgs boson production in
gluon fusion beyond NLO to be in the range from $-0.34$ to
$0.08~pb$. This falls within a more conservative estimate of
$\pm 0.40~pb$ given in \cite{Anastasiou:2016cez} on the
basis of the K-factor for the top quark mediated cross
section and the scheme dependence of the result. The above
interval can be further  reduced by evaluating  the
next-to-next-to-leading logarithmic contribution and
getting an  approximation valid beyond the threshold
region. The latter however requires the analysis of the
logarithmically enhanced corrections to the hard real
emission which currently is not available even in the LL
approximation.

\section{Conclusion}
\label{sec::5}
We have derived  the all-order next-to-leading logarithmic
approximation for the light quark loop mediated amplitude
of Higgs boson production in gluon fusion. To our knowledge
this is the first example of the subleading logarithms
resummation for a power-suppressed QCD amplitude. By using
this result an estimate of the high-order bottom  quark
contribution to the Higgs boson production cross section
has been obtained in threshold approximation. Despite a
large value of the effective expansion  parameter
$L^2\alpha_s\approx 40\alpha_s$ the corresponding
perturbative series  does converge. In NLO the
next-to-leading logarithmic approximation is in a quite
good agreement with the known  complete result. For the yet
unknown  NNLO and N$^3$LO corrections we have obtained
$-0.13~pb$ and $-0.02~pb$, respectively. With a rather
conservative assessment of accuracy of the next-to-leading
logarithmic and the threshold approximations we  give a
rough estimate of the bottom quark mediated contribution to
the total cross section of Higgs boson production in gluon
fusion beyond NLO to be in the range from $-0.34$ to
$0.08~pb$.

The actual  accuracy of the  logarithmic and threshold
approximations however is difficult to estimate and an
exact computation of quark mass effects is therefore
expected to be important in consolidating the theoretical
precision of the top-bottom interference contribution to
the inclusive Higgs cross section.  With the computation of
the complete three-loop $gg \to H$ amplitude in
Ref.~~\cite{Czakon:2020vql} and recent advances for
two-loop $pp \to H+\,jet$
amplitudes~\cite{Frellesvig:2019byn} an exact NNLO result
is within reach.

\acknowledgments
C.A is grateful to Achilleas Lazopoulos for his help in producing
numerical results with  {\tt ihixs2}~\cite{Dulat:2018rbf} and to
Nicolas Deutschmann and Armin Schweitzer for numerous discussions and
comparisons of calculations for the purposes of
Ref.~\cite{Anastasiou:2020qzk}. A.P. would like to thank Thomas
Becher and Tao Liu for useful communications. The work of A.P. is
supported in part by NSERC and Perimeter Institute for Theoretical
Physics. The authors thank the Pauli Centre of ETH Zurich for its
financial support via its visitors programme.

\vspace{4mm}

\noindent
{\bf Note added.}
In a previous version of this article the contribution of 
Eq.~(\ref{eq::coll2lg}) has been omitted. After including this
contribution the abelian part of  our result  agrees with the
results \cite{Liu:2020tzd,Liu:2020wbn,Niggetiedt:2020sbf} for
the Higgs boson two-photon decay amplitude. The numerical impact
of this contribution on the threshold cross-section at NNLO and
N$^3$LO is very small.


\begin{thebibliography}{99}

\bibitem{Cepeda:2019klc}
  M.~Cepeda {\it et al.} [HL/HE WG2 group],
  arXiv:1902.00134 [hep-ph].

\bibitem{Harlander:2009my}
  R.~V.~Harlander, H.~Mantler, S.~Marzani and K.~J.~Ozeren,
  Eur.\ Phys.\ J.\ C {\bf 66}, 359 (2010).

\bibitem{Pak:2009dg}
  A.~Pak, M.~Rogal and M.~Steinhauser,
  JHEP {\bf 1002}, 025 (2010).

\bibitem{Harlander:2009mq}
  R.~V.~Harlander and K.~J.~Ozeren,
  JHEP {\bf 0911}, 088 (2009).

\bibitem{Anastasiou:2015ema}
  C.~Anastasiou, C.~Duhr, F.~Dulat, F.~Herzog and B.~Mistlberger,
  Phys.\ Rev.\ Lett.\  {\bf 114}, 212001 (2015).

\bibitem{Anastasiou:2016cez}
  C.~Anastasiou, C.~Duhr, F.~Dulat, E.~Furlan, T.~Gehrmann, F.~Herzog, A.~Lazopoulos and B.~Mistlberger,
  JHEP {\bf 1605}, 058 (2016).



\bibitem{Mistlberger:2018etf}
  B.~Mistlberger,
  JHEP {\bf 1805}, 028 (2018).


\bibitem{Banfi:2015pju}
  A.~Banfi, F.~Caola, F.~A.~Dreyer, P.~F.~Monni, G.~P.~Salam, G.~Zanderighi and F.~Dulat,
  JHEP {\bf 1604}, 049 (2016).


\bibitem{Chen:2016zka}
  X.~Chen, J.~Cruz-Martinez, T.~Gehrmann, E.~W.~N.~Glover and M.~Jaquier,
  JHEP {\bf 1610}, 066 (2016).


\bibitem{Caola:2015wna}
  F.~Caola, K.~Melnikov and M.~Schulze,
  Phys.\ Rev.\ D {\bf 92},  074032 (2015).


\bibitem{Chen:2014gva}
  X.~Chen, T.~Gehrmann, E.~W.~N.~Glover and M.~Jaquier,
  Phys.\ Lett.\ B {\bf 740}, 147 (2015).


\bibitem{Dulat:2017prg}
  F.~Dulat, B.~Mistlberger and A.~Pelloni,
  JHEP {\bf 1801}, 145 (2018).


\bibitem{Dulat:2018bfe}
  F.~Dulat, B.~Mistlberger and A.~Pelloni,
  Phys.\ Rev.\ D {\bf 99},  034004 (2019).


\bibitem{Cieri:2018oms}
  L.~Cieri, X.~Chen, T.~Gehrmann, E.~W.~N.~Glover and A.~Huss,
  JHEP {\bf 1902}, 096 (2019).




\bibitem{Graudenz:1992pv}
  D.~Graudenz, M.~Spira and P.~M.~Zerwas,
  Phys.\ Rev.\ Lett.\  {\bf 70}, 1372 (1993).


\bibitem{Djouadi:1991tka}
  A.~Djouadi, M.~Spira and P.~M.~Zerwas,
  Phys.\ Lett.\ B {\bf 264}, 440 (1991).


\bibitem{Spira:1995rr}
  M.~Spira, A.~Djouadi, D.~Graudenz and P.~M.~Zerwas,
  Nucl.\ Phys.\ B {\bf 453}, 17 (1995).


\bibitem{Harlander:2005rq}
  R.~Harlander and P.~Kant,
  JHEP {\bf 0512}, 015 (2005).


\bibitem{Aglietti:2006tp}
  U.~Aglietti, R.~Bonciani, G.~Degrassi and A.~Vicini,
  JHEP {\bf 0701}, 021 (2007).


\bibitem{Bonciani:2007ex}
  R.~Bonciani, G.~Degrassi and A.~Vicini,
  JHEP {\bf 0711}, 095 (2007).


\bibitem{Anastasiou:2006hc}
  C.~Anastasiou, S.~Beerli, S.~Bucherer, A.~Daleo and Z.~Kunszt,
  JHEP {\bf 0701}, 082 (2007).


\bibitem{Anastasiou:2009kn}
  C.~Anastasiou, S.~Bucherer and Z.~Kunszt,
  JHEP {\bf 0910}, 068 (2009).




\bibitem{Melnikov:2016qoc}
  K.~Melnikov, L.~Tancredi and C.~Wever,
  JHEP {\bf 1611}, 104 (2016).



\bibitem{Melnikov:2017pgf}
  K.~Melnikov, L.~Tancredi and C.~Wever,
  Phys.\ Rev.\ D {\bf 95},  054012 (2017).


\bibitem{Lindert:2017pky}
  J.~M.~Lindert, K.~Melnikov, L.~Tancredi and C.~Wever,
  Phys.\ Rev.\ Lett.\  {\bf 118},  252002 (2017).



\bibitem{Liu:2017vkm}
  T.~Liu and A.~A.~Penin,
  Phys.\ Rev.\ Lett.\  {\bf 119},  262001 (2017).





\bibitem{Liu:2018czl}
  T.~Liu and A.~Penin,
  JHEP {\bf 1811}, 158 (2018).


\bibitem{Melnikov:2016emg}
  K.~Melnikov and A.~Penin,
  JHEP {\bf 1605}, 172 (2016).


\bibitem{Sudakov:1954sw}
  V.~V.~Sudakov,
  Sov.\ Phys.\ JETP {\bf 3}, 65 (1956)
  [Zh.\ Eksp.\ Teor.\ Fiz.\  {\bf 30}, 87 (1956)].


\bibitem{Frenkel:1976bj}
  J.~Frenkel and J.~C.~Taylor,
  Nucl.\ Phys.\ B {\bf 116}, 185 (1976).

\bibitem{Smilga:1979uj}
  A.~V.~Smilga,
  Nucl.\ Phys.\ B {\bf 161}, 449 (1979).

\bibitem{Mueller:1979ih}
  A.~H.~Mueller,
  Phys.\ Rev.\ D {\bf 20}, 2037 (1979).

\bibitem{Collins:1980ih}
  J.~C.~Collins,
  Phys.\ Rev.\ D {\bf 22}, 1478 (1980).

\bibitem{Sen:1981sd}
  A.~Sen,
  Phys.\ Rev.\ D {\bf 24}, 3281 (1981).

\bibitem{Sterman:1986aj}
  G.~F.~Sterman,
  Nucl.\ Phys.\ B {\bf 281}, 310 (1987).

\bibitem{Korchemsky:1988pn}
  G.~P.~Korchemsky,
  Phys.\ Lett.\ B {\bf 217}, 330 (1989).

\bibitem{Korchemsky:1988hd}
  G.~P.~Korchemsky,
  Phys.\ Lett.\ B {\bf 220}, 629 (1989).



\bibitem{Kuhn:1999nn}
  J.~H.~Kuhn, A.~A.~Penin and V.~A.~Smirnov,
  Eur.\ Phys.\ J.\ C {\bf 17}, 97 (2000).


\bibitem{Kuhn:2001hz}
  J.~H.~Kuhn, S.~Moch, A.~A.~Penin and V.~A.~Smirnov,
  Nucl.\ Phys.\ B {\bf 616}, 286 (2001),
  Erratum: [Nucl.\ Phys.\ B {\bf 648}, 455 (2003)].

\bibitem{Feucht:2004rp}
  B.~Feucht, J.~H.~Kuhn, A.~A.~Penin and V.~A.~Smirnov,
  Phys.\ Rev.\ Lett.\  {\bf 93}, 101802 (2004).


\bibitem{Jantzen:2005az}
  B.~Jantzen, J.~H.~K\"uhn, A.~A.~Penin and V.~A.~Smirnov,
  Nucl.\ Phys.\ B {\bf 731}, 188 (2005).


\bibitem{Penin:2005kf}
  A.~A.~Penin,
  Phys.\ Rev.\ Lett.\  {\bf 95}, 010408 (2005).

\bibitem{Penin:2005eh}
  A.~A.~Penin,
  Nucl.\ Phys.\ B {\bf 734},  185 (2006).


\bibitem{Bonciani:2007eh}
  R.~Bonciani, A.~Ferroglia, and A.~A.~Penin,
  Phys.\ Rev.\ Lett.\  {\bf 100},  131601 (2008).

\bibitem{Bonciani:2008ep}
  R.~Bonciani, A.~Ferroglia, and A.~A.~Penin,
  JHEP {\bf 0802},  080 (2008).

\bibitem{Kuhn:2007ca}
  J.~H.~K\"uhn, F.~Metzler and A.~A.~Penin,
  Nucl.\ Phys.\ B {\bf 795}, 277 (2008).

\bibitem{Kuhn:2011mh}
  J.~H.~K\"uhn, F.~Metzler, A.~A.~Penin, and S.~Uccirati,
  JHEP {\bf 1106}, 143 (2011).


\bibitem{Penin:2011aa}
  A.~A.~Penin and G.~Ryan,
  JHEP {\bf 1111},  081 (2011).

\bibitem{Gorshkov:1966ht}
  V.~G.~Gorshkov, V.~N.~Gribov, L.~N.~Lipatov and G.~V.~Frolov,
  Sov.\ J.\ Nucl.\ Phys.\  {\bf 6}, 95 (1968)
  [Yad.\ Fiz.\  {\bf 6}, 129 (1967)].


\bibitem{Kotsky:1997rq}
  M.~I.~Kotsky and O.~I.~Yakovlev,
  Phys.\ Lett.\ B {\bf 418}, 335 (1998).

\bibitem{Akhoury:2001mz}
  R.~Akhoury, H.~Wang and O.~I.~Yakovlev,
  Phys.\ Rev.\ D {\bf 64}, 113008 (2001).

\bibitem{Ferroglia:2009ep}
  A.~Ferroglia, M.~Neubert, B.~D.~Pecjak and L.~L.~Yang,
  Phys.\ Rev.\ Lett.\  {\bf 103} (2009) 201601.


\bibitem{Laenen:2010uz}
  E.~Laenen, L.~Magnea, G.~Stavenga and C.~D.~White,
  JHEP {\bf 1101}, 141 (2011).

\bibitem{Banfi:2013eda}
  A.~Banfi, P.~F.~Monni and G.~Zanderighi,
  JHEP {\bf 1401}, 097 (2014).

\bibitem{Becher:2013iya}
  T.~Becher and G.~Bell,
  Phys.\ Rev.\ Lett.\  {\bf 112}, 182002 (2014).

\bibitem{deFlorian:2014vta}
  D.~de Florian, J.~Mazzitelli, S.~Moch and A.~Vogt,
  JHEP {\bf 1410}, 176 (2014).

\bibitem{Anastasiou:2014lda}
  C.~Anastasiou, C.~Duhr, F.~Dulat, E.~Furlan, T.~Gehrmann, F.~Herzog and
B.~Mistlberger,
  JHEP {\bf 1503}, 091 (2015)

\bibitem{Penin:2014msa}
  A.~A.~Penin,
  Phys.\ Lett.\ B {\bf 745}, 69 (2015), Erratum: [Phys.\ Lett.\ B {\bf 771}, 633
(2017)].

\bibitem{Almasy:2015dyv}
  A.~A.~Almasy, N.~A.~Lo Presti and A.~Vogt,
  JHEP {\bf 1601}, 028 (2016).


\bibitem{Penin:2016wiw}
  A.~A.~Penin and N.~Zerf,
  Phys.\ Lett.\ B {\bf 760}, 816 (2016),  Erratum: [Phys.\ Lett.\ B {\bf 771},
637 (2017)].


\bibitem{Bonocore:2016awd}
  D.~Bonocore, E.~Laenen, L.~Magnea, L.~Vernazza and C.~D.~White,
  JHEP {\bf 1612}, 121 (2016).

\bibitem{Boughezal:2016zws}
  R.~Boughezal, X.~Liu and F.~Petriello,
  JHEP {\bf 1703}, 160 (2017).


\bibitem{Moult:2017rpl}
  I.~Moult, I.~W.~Stewart and G.~Vita,
  JHEP {\bf 1707}, 067 (2017).

\bibitem{Liu:2017axv}
  T.~Liu, A.~A.~Penin and N.~Zerf,
  Phys.\ Lett.\ B {\bf 771}, 492 (2017).


\bibitem{Beneke:2017ztn}
  M.~Beneke, M.~Garny, R.~Szafron and J.~Wang,
  JHEP {\bf 1803}, 001 (2018).

\bibitem{Boughezal:2018mvf}
  R.~Boughezal, A.~Isgr\'o and F.~Petriello,
  Phys.\ Rev.\ D {\bf 97},  076006 (2018).

\bibitem{Bruser:2018jnc}
  R.~Br\"user, S.~Caron-Huot and J.~M.~Henn,
  JHEP {\bf 1804}, 047 (2018).

  \bibitem{Moult:2018jjd}
  I.~Moult, I.~W.~Stewart, G.~Vita and H.~X.~Zhu,
  JHEP {\bf 1808}, 013 (2018).

\bibitem{Ebert:2018lzn}
  M.~A.~Ebert, I.~Moult, I.~W.~Stewart, F.~J.~Tackmann, G.~Vita and H.~X.~Zhu,
  JHEP {\bf 1812}, 084 (2018).

\bibitem{Alte:2018nbn}
  S.~Alte, M.~K\"onig and M.~Neubert,
  JHEP {\bf 1808}, 095 (2018).

\bibitem{Beneke:2018rbh}
  M.~Beneke, M.~Garny, R.~Szafron and J.~Wang,
  JHEP {\bf 1811}, 112 (2018).

\bibitem{Beneke:2018gvs}
  M.~Beneke, A.~Broggio, M.~Garny, S.~Jaskiewicz, R.~Szafron, L.~Vernazza and J.~Wang,
  JHEP {\bf 1903}, 043 (2019).

\bibitem{Engel:2018fsb}
  T.~Engel, C.~Gnendiger, A.~Signer and Y.~Ulrich,
  JHEP {\bf 1902}, 118 (2019).

\bibitem{Ebert:2018gsn}
  M.~A.~Ebert, I.~Moult, I.~W.~Stewart, F.~J.~Tackmann, G.~Vita and H.~X.~Zhu,
  JHEP {\bf 1904}, 123 (2019).

\bibitem{Penin:2019xql}
  A.~A.~Penin,
  JHEP {\bf 2004}, 156 (2020).

\bibitem{Beneke:2019mua}
  M.~Beneke, M.~Garny, S.~Jaskiewicz, R.~Szafron, L.~Vernazza and J.~Wang,
  JHEP {\bf 2001}, 094 (2020).

\bibitem{Liu:2019oav}
  Z.~L.~Liu and M.~Neubert,
  arXiv:1912.08818 [hep-ph].

\bibitem{Wang:2019mym}
  J.~Wang,
  arXiv:1912.09920 [hep-ph].


\bibitem{Beneke:1997zp}
  M.~Beneke and V.~A.~Smirnov,
  Nucl.\ Phys.\ B {\bf 522}, 321 (1998).

\bibitem{Smirnov:1997gx}
  V.~A.~Smirnov,
  Phys.\ Lett.\ B {\bf 404}, 101 (1997).

\bibitem{Smirnov:2002pj}
  V.~A.~Smirnov,
  {\it Applied asymptotic expansions in momenta and masses},
  Springer Tracts Mod.\ Phys.\  {\bf 177 }, 1 (2002).


\bibitem{Yennie:1961ad}
  D.~R.~Yennie, S.~C.~Frautschi and H.~Suura,
  Annals Phys.\  {\bf 13}, 379 (1961).


\bibitem{Anastasiou:2020qzk}
  C.~Anastasiou, N.~Deutschmann and A.~Schweitzer,
  arXiv:2001.06295 [hep-ph].

\bibitem{Catani:1998bh}
  S.~Catani,
  Phys.\ Lett.\ B {\bf 427}, 161 (1998).

\bibitem{Frenkel:1984pz}
  J.~Frenkel and J.~C.~Taylor,
  Nucl.\ Phys.\ B {\bf 246}, 231 (1984).



\bibitem{Harlander:2019ioe}
  R.~V.~Harlander, M.~Prausa and J.~Usovitsch,
  JHEP {\bf 1910}, 148 (2019).


\bibitem{Czakon:2020vql}
  M.~Czakon and M.~Niggetiedt,
  arXiv:2001.03008 [hep-ph].

\bibitem{Davies:2019nhm}
  J.~Davies, R.~Gröber, A.~Maier, T.~Rauh and M.~Steinhauser,
  Phys.\ Rev.\ D {\bf 100}, 034017 (2019).


\bibitem{Chetyrkin:1997un}
  K.~G.~Chetyrkin, B.~A.~Kniehl and M.~Steinhauser,
  Nucl.\ Phys.\ B {\bf 510}, 61 (1998).

\bibitem{Schroder:2005hy}
  Y.~Schroder and M.~Steinhauser,
  JHEP {\bf 0601}, 051 (2006).

\bibitem{Chetyrkin:2005ia}
  K.~G.~Chetyrkin, J.~H.~Kuhn and C.~Sturm,
  Nucl.\ Phys.\ B {\bf 744}, 121 (2006).


\bibitem{Catani:2001ic}
  S.~Catani, D.~de Florian and M.~Grazzini,
  JHEP {\bf 0105}, 025 (2001).

\bibitem{Harlander:2001is}
  R.~V.~Harlander and W.~B.~Kilgore,
  Phys.\ Rev.\ D {\bf 64}, 013015 (2001).


\bibitem{Anastasiou:2014vaa}
  C.~Anastasiou, C.~Duhr, F.~Dulat, E.~Furlan, T.~Gehrmann, F.~Herzog and B.~Mistlberger,
  Phys.\ Lett.\ B {\bf 737}, 325 (2014).


\bibitem{Dulat:2018rbf}
  F.~Dulat, A.~Lazopoulos and B.~Mistlberger,
  Comput.\ Phys.\ Commun.\  {\bf 233}, 243 (2018).


\bibitem{Frellesvig:2019byn}
  H.~Frellesvig, M.~Hidding, L.~Maestri, F.~Moriello and G.~Salvatori,
  arXiv:1911.06308 [hep-ph].


\bibitem{Liu:2020tzd}
  Z.~L.~Liu, B.~Mecaj, M.~Neubert and X.~Wang,
  arXiv:2009.04456 [hep-ph].

\bibitem{Liu:2020wbn}
  Z.~L.~Liu, B.~Mecaj, M.~Neubert and X.~Wang,
  arXiv:2009.06779 [hep-ph].

\bibitem{Niggetiedt:2020sbf}
  M.~Niggetiedt,
  arXiv:2009.10556 [hep-ph].

\end{thebibliography}
\end{document}